\documentclass[12pt]{article}
\usepackage{epsf}
\usepackage{epsfig}
\usepackage{amsfonts}
\usepackage{wasysym}

\newcommand{\bea}{\begin{eqnarray}}
\newcommand{\eea}{\end{eqnarray}}
\newcommand{\ba}{\begin{array}}
\newcommand{\ea}{\end{array}}

\def\nn{\nonumber}

\renewcommand{\b}{{\mathbb B}}
\renewcommand{\c}{{\mathbb C}}

\newcommand{\U}{{\bf U}}
\newcommand{\V}{{\bf V}}
\newcommand{\N}{{\bf N}}
\renewcommand{\S}{{\bf S}}
\def\gappeq{\mathrel{\rlap {\raise.5ex\hbox{$>$}}
{\lower.5ex\hbox{$\sim$}}}}
\def\lappeq{\mathrel{\rlap{\raise.5ex\hbox{$<$}}
{\lower.5ex\hbox{$\sim$}}}}

\renewcommand{\arraystretch}{1.3}
\newcommand{\lsim}{\raisebox{-0.13cm}{~\shortstack{$<$ \\[-0.07cm] $\sim$}}~}

\begin{document}
\topmargin -1.0cm
\oddsidemargin -0.4cm
\evensidemargin -0.4cm
\pagestyle{empty}
\begin{flushright}
UG-FT-137/02\\
CAFPE-7/02\\
hep-ph/0207328\\
rev. November 2002
\end{flushright}
\vspace*{5mm}
\begin{center}

{\Large\bf Lepton Flavor Violation  in \boldmath$Z$ and Lepton Decays}

{\Large\bf in Supersymmetric Models}

\vspace{1.4cm}
{\sc J.~I. Illana and M. Masip}\\
\vspace{.8cm}
{\it Centro Andaluz de F\'\i sica de Part\'\i culas Elementales (CAFPE) and}\\
{\it Departamento de F\'\i sica Te\'orica y del Cosmos}\\
{\it Universidad de Granada}\\
{\it E-18071 Granada, Spain}\\

\end{center}
\vspace{1.4cm}
\begin{abstract}

The observation of charged lepton flavor non--conservation
would be a clear signature of physics beyond the Standard Model.
In particular, supersymmetric (SUSY) models introduce mixings
in the sneutrino and the charged slepton sectors which could
imply flavor--changing processes at rates accessible
to upcoming experiments. In this paper we analyze the possibility
to observe $Z\rightarrow \ell_I \ell_J$ in the GigaZ option of TESLA
at DESY.
We show that although models with SUSY masses above the current 
limits could predict a branching ratio ${\rm BR}(Z\rightarrow \mu e)$ 
accessible to the experiment, they would imply an unobserved rate of 
$\mu\rightarrow e \gamma$ and thus are excluded. 
In models with a small mixing angle between the first and the third
(or the second and the third) slepton families GigaZ could observe
$Z\rightarrow \tau\mu$ (or $Z\rightarrow \tau e$) consistently 
with present bounds on $\ell_J\rightarrow \ell_I \gamma$.
In contrast, if the mixing angles between the three slepton 
families are large the bounds from $\mu\rightarrow e \gamma$ 
push these processes below the reach of GigaZ. 
We show that in this case the masses of the three
slepton families must be strongly degenerated (with mass 
differences of order $10^{-3}$).
We update the limits on
the slepton mass insertions $\delta_{LL,RR,LR}$ and 
discuss the correlation between
flavor changing and $g_\mu-2$ in SUSY models.
\end{abstract}

\centerline{PACS numbers: 12.60.Jv, 13.35.-r, 13.38.Dg}

\vfill

\eject

\pagestyle{empty}
\setcounter{page}{1}
\setcounter{footnote}{0}
\pagestyle{plain}


\section{Introduction}

Lepton flavor violation (LFV) has been searched in 
several experiments. The current status 
in $\mu$ and $\tau$ decays is
\begin{eqnarray}
{\rm BR}(\mu\rightarrow e \gamma)&<&1.2\times 10^{-11} 
\quad\cite{Brooks:1999pu} \;, \nonumber \\
{\rm BR}(\tau\rightarrow e \gamma)&<&2.7\times 10^{-6} 
\quad\cite{Edwards:1996te} \;, \nonumber \\
{\rm BR}(\tau\rightarrow \mu \gamma)&<&1.1\times 10^{-6} 
\quad\cite{Ahmed:1999gh} \;,
\label{br1}
\end{eqnarray}
and
\begin{eqnarray}
{\rm BR}(\mu\rightarrow 3e)&<&1.0\times 10^{-12} 
\quad\cite{Bellgardt:1987du} \;, \nonumber \\
{\rm BR}(\tau\rightarrow 3e)&<&2.9\times 10^{-6} 
\quad\cite{Bliss:1997iq} \;, \nonumber \\
{\rm BR}(\tau\rightarrow 3\mu)&<&1.9\times 10^{-6} 
\quad\cite{Bliss:1997iq} \;.
\label{br2}
\end{eqnarray}
In $Z$ decays we have
\begin{eqnarray}
{\rm BR}(Z\rightarrow \mu e)&<&1.7\times 10^{-6} 
\quad\cite{Akers:1995gz} \;, \nonumber \\
{\rm BR}(Z\rightarrow \tau e)&<&9.8\times 10^{-6} 
\quad\cite{Akers:1995gz} \;, \nonumber \\
{\rm BR}(Z\rightarrow \tau \mu)&<&1.2\times 10^{-5} 
\quad\cite{Abreu:1996mj} \;.
\label{br3}
\end{eqnarray}
These observations are obviously in agreement with the Standard
Model (SM), where lepton flavor number is (perturbatively)
conserved.

On the other hand, neutrino oscillations are a first evidence 
of LFV. Small neutrino masses and mixings of order one suggest the 
existence of a new scale around $10^{12}$ GeV \cite{seesaw}. Massive
neutrinos could be 
naturally accommodated within the SM (the so called
$\nu$SM). The contributions from the light neutrino sector to 
other LFV processes, however, would be very small:  
BR$(\ell_J\rightarrow \ell_I \gamma)\lsim 10^{-48}$ and 
BR$(Z\rightarrow \ell_I \ell_J)\lsim 10^{-54}$ \cite{Illana:2000ic}.
In consequence, any experimental signature of LFV in 
the charged sector would be a clear signature of nonstandard physics.

In this paper we will study the implications of 
supersymmetry (SUSY) on $Z\rightarrow \ell_I \ell_J$.\footnote{
A recent work on the flavor--changing decays
$Z\to d_I d_J$ in 2HDMs and SUSY has been presented in 
\cite{Atwood:2002ke}.}
The GigaZ option of the TESLA Linear Collider project 
\cite{Aguilar-Saavedra:2001rg} could reduce the 
LEP bounds down to \cite{Wilson:1998bb}
\begin{eqnarray}
{\rm BR}(Z\rightarrow \mu e)&<&2.0\times 10^{-9} \nonumber \;, \\
{\rm BR}(Z\rightarrow \tau e)&<&\kappa\times6.5\times 10^{-8}\;, \nonumber \\
{\rm BR}(Z\rightarrow \tau \mu)&<&\kappa\times2.2\times 10^{-8} \;,
\label{br4}
\end{eqnarray}
with $\kappa=0.2-1.0$.
We will here explore the possibility that SUSY provides a signal
accessible to GigaZ in consistency with current bounds
from BR$(\ell_J\rightarrow \ell_I \gamma)$. Note that 
in SUSY models the branching ratio
${\rm BR}(\ell_J\rightarrow 3 \ell_I) \approx \alpha_{em}
{\rm BR}(\ell_J\rightarrow \ell_I \gamma)$ 
will place weaker bounds
on SUSY parameters (see Eqs.~(1,2))
The conversion rate $\mu\rightarrow e$ on Ti 
gives also weaker bounds at current experiments, although this may
change in the future (see \cite{Hisano:2002zu} for a recent review).

We will concentrate on the minimal SUSY 
extension of the SM (MSSM) with R-parity and general
soft SUSY--breaking terms. 
Related works on LFV in $Z$ decays in SUSY models study 
the MSSM \cite{Levine:1987fv} and 
a left--right SUSY model \cite{Frank:1996xt}. 
Several groups have analyzed other LFV processes 
in SUSY grand unified models
with massive neutrinos (motivated by the atmospheric and solar 
neutrino anomalies \cite{susygut}),
or with R--parity violation \cite{Carvalho:2002bq}. 
There are also studies \cite{Delepine:2001di,Frank:2001dc} 
relating LFV $Z$ decays with other processes.
Direct signals
of lepton flavor non--conservation in slepton production at the 
LHC \cite{Hinchliffe:2000np} and at future $e^+e^-$ or
$\mu^+\mu^-$ colliders \cite{directlc} have been also explored.

Other works on LFV $Z$ decays in alternative models include
the SM with massive Dirac or Majorana neutrinos 
\cite{Illana:2000ic,Ilakovac:1994kj}, left--right symmetric models 
\cite{Bernabeu:1995ph},
models with a heavy $Z'$ boson \cite{Langacker:2000ju},
two Higgs doublet models (2HDMs) \cite{Iltan:2001au} or
technicolor \cite{Yue:2002pk}.


\section{Calculation}

The most general vertex $V \bar\ell_I\ell_J$ coupling a 
(lepton) fermion current to a vector
boson can be parametrized in terms of four form factors:
\bea
{\cal M}=i \varepsilon^\mu_V \bar u_{\ell_I}(p_2)
\left[\gamma_\mu(F_V-F_A\gamma_5)+(iF_M+F_E\gamma_5)\sigma_{\mu\nu}q^\nu
      \right] u_{\ell_J}(p_1)\;,
\label{M0}
\eea
where $\varepsilon_V$ is the polarization vector ($\varepsilon_V\cdot q=0$) 
and $q=p_2-p_1$ is the momentum transfer. For
an on--shell (massless) photon $F_A=0$, and, in addition, if
$m_I\not=m_J$ then $F_V=0$. This implies that the flavor--changing
process $\ell_J \rightarrow \ell_I \gamma$
is determined by (chirality flipping) dipole transitions only. 
In contrast, all form factors contribute to the decay of a $Z$ boson:
\bea
{\rm BR}(Z\to \ell_I \ell_J)&\equiv &{\rm BR}(Z\to \bar \ell_I \ell_J) + 
{\rm BR}(Z\to \ell_I \bar \ell_J)\nn\\
&=&\frac{\alpha_W^3M_Z}{48\pi^2\Gamma_Z}
\left[|f_L|^2+|f_R|^2+\frac{1}{c^2_W}(|f_M|^2+|f_E|^2)\right],
\label{BR}
\eea
with $\alpha_W=g^2/(4\pi)$, $f_{L,R}=f_V\pm f_A$, $f_{V,A}\equiv 
-16\pi^2g^{-3} 
F_{V,A}$ and $f_{M,E}\equiv -16M_W\pi^2g^{-3} F_{M,E}$.
We calculate (see Appendices A and B for details) 
these branching ratios in the MSSM. 

Let us consider the case with two 
lepton families. Since SUSY is broken, fermion and scalar mass
matrices will be diagonalized by different rotations in flavor
space. After the diagonalization of the fermion sector we are left
with a $2\times 2$ scalar matrix with 3 arbitrary parameters.
We will assume that the rotation that 
diagonalizes the 
scalar matrix is maximal, $\theta=\pi/4$ (i.e. we assume no
alignment between fermion and scalar fields; the amplitudes that
we will calculate are proportional to $\sin2\theta$). 
Our choice corresponds to a mass matrix with 
identical diagonal terms. Taking 
\bea 
{\bf m}^2=\tilde m^2 \pmatrix{ \sqrt{1+\delta^2} & \delta \cr
\delta & \sqrt{1+\delta^2} \cr}\;,
\eea
the two mass eigenvalues are
\bea
\tilde m_{1,2}^2=\tilde m^2 (\sqrt{1+\delta^2}\mp \delta)\;.
\eea
In this parametrization $\tilde m^2=\tilde m_1 \tilde m_2$ 
characterizes the SUSY--breaking scale
and $\delta=(\tilde m_2^2-\tilde m_1^2)/(2\tilde m^2)$ the mass 
splitting between the two families.
$\delta$ is also responsible for any flavor--changing
process: 
$\delta=0$ corresponds to the flavor--conserving case, $\delta\ll 1$
can be treated as a non--diagonal mass insertion, and 
$\delta\rightarrow \infty$
gives $\tilde m_2^2\rightarrow \infty$ (a decoupled second family).
The last case implies a 
maximum flavor--changing rate \cite{Levine:1987fv,Frank:1996xt}.

To analyze the general case with three lepton families we will 
consider two scenarios. 
First, we will follow the usual approach \cite{Masiero}
where the influence of a nondiagonal term $\tilde\delta^{IJ}$
is calculated putting the rest to zero. This implies that
the slepton family $\ell_{K\not=I,J}$ does not mix with $\ell_I$ 
and $\ell_J$
($\theta_{IK}=\theta_{JK}=0$), which reduces the problem to 
the two family case discussed above. This approximation 
is only justified if the off--diagonal terms satisfy
$\tilde\delta^{IK} \tilde\delta^{KJ} < \tilde\delta^{IJ}$ or, 
in terms of mixing
angles and mass differences, if
\bea \left( {\tilde m_I^2-\tilde m_K^2\over \tilde m_I^2+\tilde m_K^2}
\sin \theta_{IK} \right) 
\left( {\tilde m_K^2-\tilde m_J^2\over \tilde m_K^2+\tilde m_J^2}
\sin \theta_{JK}\right) \lsim
{\tilde m_I^2-\tilde m_J^2\over \tilde m_I^2+\tilde m_J^2}
\sin \theta_{IJ}\;. 
\eea
We will then discuss a second scenario with maximal mixing 
between the three slepton families: $\theta_{12}=\theta_{23}=
\theta_{13}=\pi/4$. Large mixings are suggested 
by the observation of solar and atmospheric neutrino 
oscilations (note, however, that the non observation of 
$\nu_e-\nu_\tau$ oscilations in CHOOZ \cite{Apollonio:1999ae}
could suggest 
$\theta_{13}\lsim 0.1$). We will use $\delta^{IJ}$ to 
parametrize the mass difference between $\ell_I$ and $\ell_J$:
$\delta^{IJ}\equiv (\tilde m_J^2-\tilde m_I^2) / (2 \tilde m_1^2)$, 
where $\tilde m_1^2$ is the mass of the lightest slepton family.

The relevant parameters for the calculation will then be 
the masses and mixings of charginos and neutralinos, the masses 
of the six (`left' and `right' handed) charged sleptons, 
and the masses of the three sneutrinos. 
When we evaluate the contribution of each 
$\delta^{IJ}$ to $Z\rightarrow \ell_I \ell_J$ and 
$\ell_J\rightarrow \ell_I \gamma$
setting all the other to zero we will have three independent
parameters $\delta^{\tilde \nu\,IJ}_{LL}$ in the sneutrino sector 
and nine $\delta^{\tilde \ell\,IJ}_{LL}$,
$\delta^{\tilde \ell\,IJ}_{RR}$ and $\delta^{\tilde \ell\,IJ}_{LR}$ 
in the charged slepton sector.
In the case of maximal mixing between the three slepton families there 
will be two independent mass differences $\delta^{\tilde \nu\,IJ}_{LL}$
and four $\delta^{\tilde \ell\,IJ}_{LL}$, $\delta^{\tilde \ell\,IJ}_{RR}$
(note that in this case $\delta^{23}=\delta^{13}-\delta^{12}$).

In our analysis we will not assume any (grand unification)
relation between slepton masses.
For each non--zero choice of $\delta^{IJ}$ it is straightforward 
to obtain and diagonalize the mass matrix that corresponds to a maximal
rotation angle (see Appendix B).
Our results should coincide with the ones
obtained in the limit of small mass difference using the mass insertion 
method, but they are also valid for any large value of $\delta^{IJ}$.

\begin{figure}
\centerline{\epsfig{file=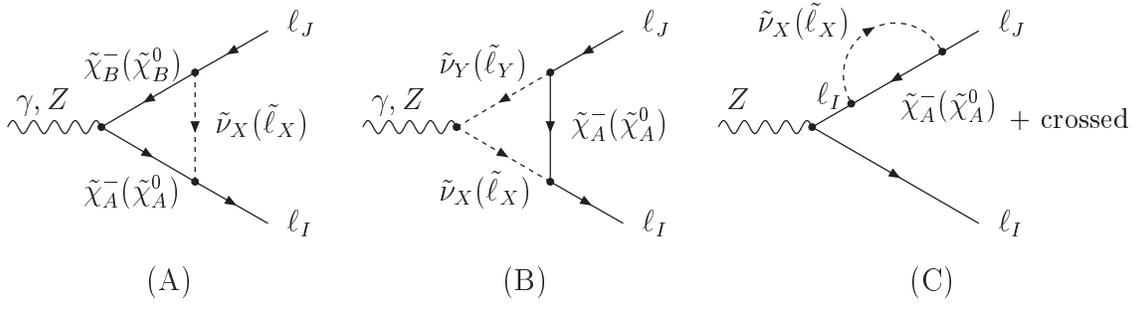,width=0.9\linewidth}}
\caption{SUSY contributions to the LFV processes $Z\to\ell_I\ell_J$ and
$\ell_J\to\ell_I\gamma$.
\label{fig1}}
\end{figure}

The process $Z\rightarrow \ell_I \ell_J$ goes through the diagrams in 
Fig.~\ref{fig1}.
Box diagrams mediating $e^+e^-\to\ell_I\ell_J$ introduce a small 
correction of order $\Gamma_Z/M_Z\approx\alpha$.
Analogous diagrams describe $\ell_J\rightarrow \ell_I \gamma$. The 
inclusion of the contributions of the third type is essential to cancel
ultraviolet divergences (they are related to counterterms by Ward 
identities). Diagrams with neutralinos in (A) or sneutrinos in (B) do not
couple to the photon. The diagrams of type (C) do not give dipole 
contributions.


\begin{figure}
\begin{center}
\begin{tabular}{cc}
\epsfig{file=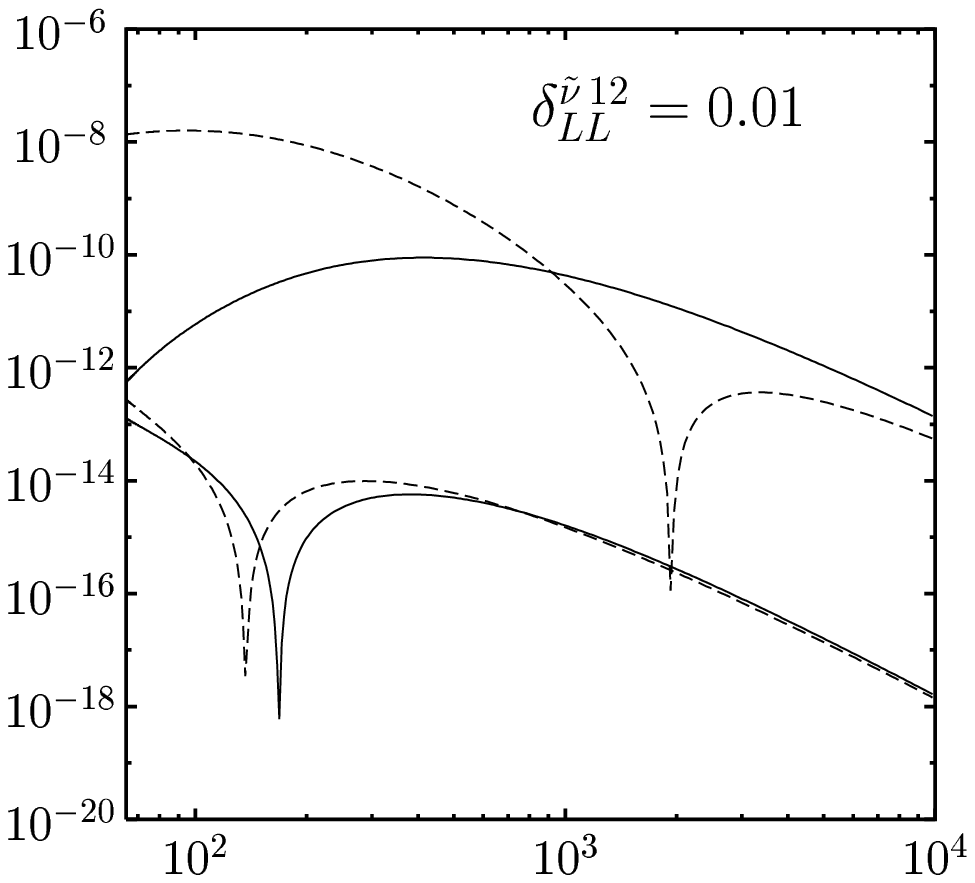,width=0.45\linewidth} &
\epsfig{file=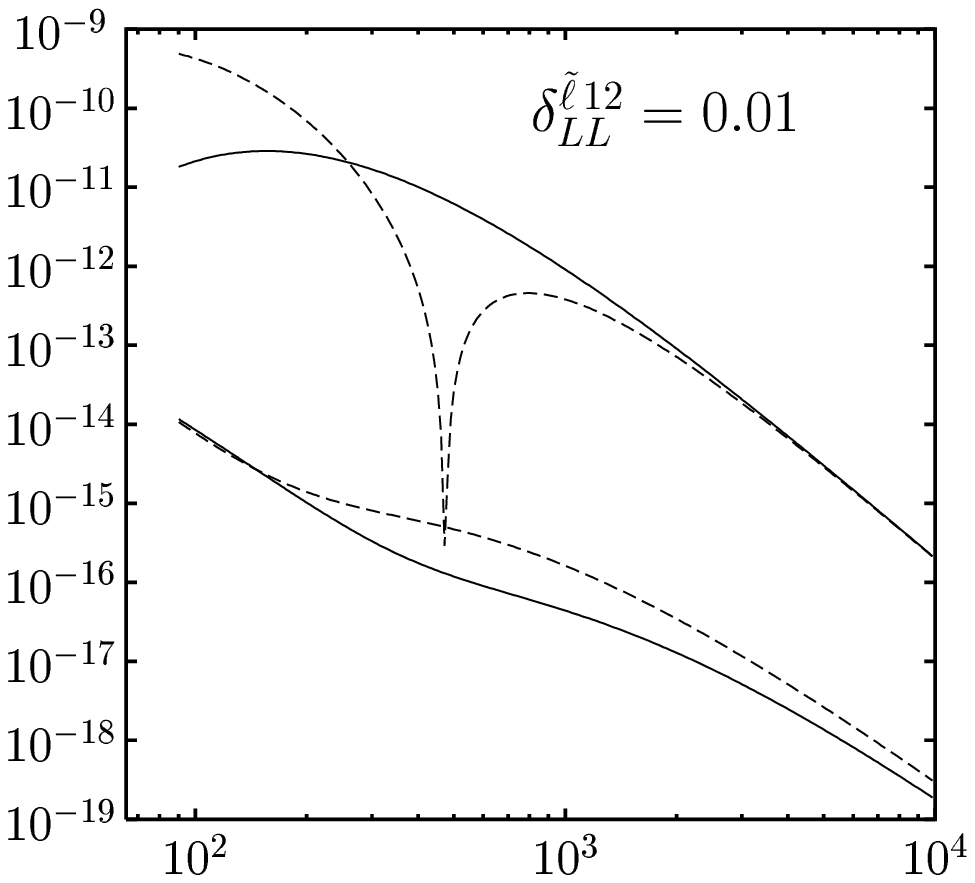,width=0.45\linewidth} \\
\epsfig{file=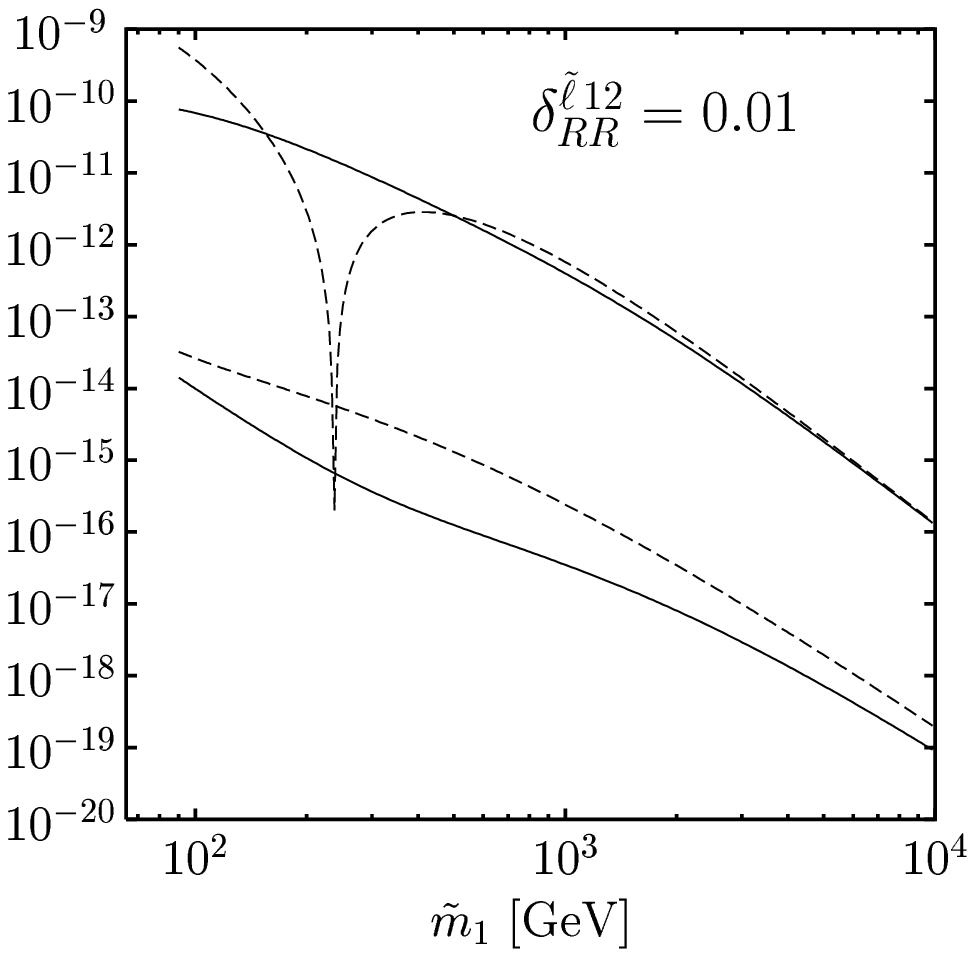,width=0.45\linewidth} &
\epsfig{file=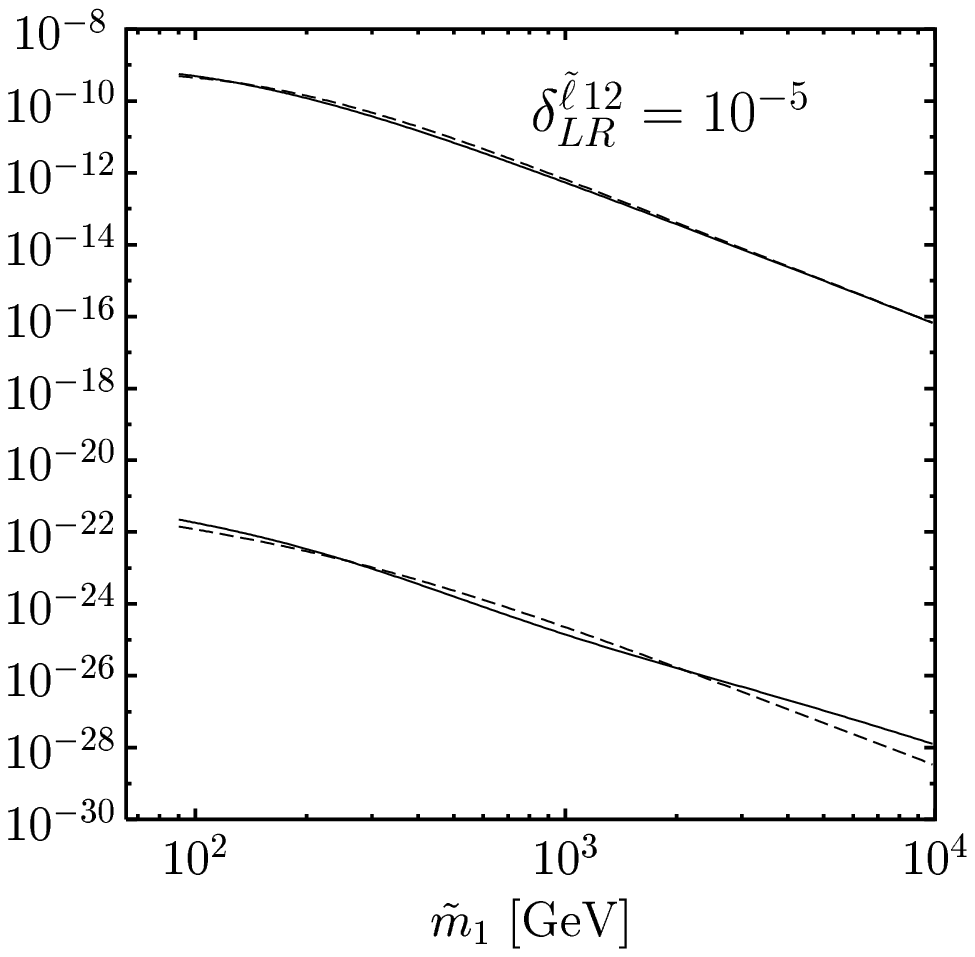,width=0.45\linewidth}
\end{tabular}
\end{center}
\caption{BR$(Z\to\mu e)$ (lower curves) and BR$(\mu\to e\gamma)$
(upper curves) as a function of the lightest scalar mass $\tilde m_1$ for 
$\tan\beta=2$ and the different $\delta^{12}$. 
Solid lines correspond to $M_2=150$ GeV and $\mu=-500$ GeV and
dashed lines to $M_2=\mu=150$ GeV.
\label{fig2}}
\end{figure}

\begin{table}[ht]
\caption{Approximate lower bounds on SUSY mass parameters based on 
\cite{Groom:in}. Note that,
for negligible scalar trilinears,
$m_{\tilde\nu}^2=m_{\tilde\ell_L}^2+M^2_Zc^2_W\cos2\beta$,
and the bounds on $m_{\tilde\nu}$ and $m_{\tilde\ell_L}$ are 
correlated. For instance:
$m_{\tilde\nu}> 65\ (40)\ \mbox{GeV for}\ \tan\beta=2\ (50)$.
\label{tab1}}
\begin{center}
\begin{tabular}{|c|l|}
\hline
lightest slepton $(\tilde m_1)$  & $m_{\tilde\nu} \ge 45$~GeV \\
 				 & $m_{\tilde\ell_{L,R}} \ge 90$~GeV
\\
\hline
lightest chargino & $m_{\tilde\chi^+_1} \ge 75$~GeV, 
					if $m_{\tilde\nu}>m_{\chi^+_1}$ \\
                   & $m_{\tilde\chi^+_1} \ge 45$~GeV, otherwise 
\\
\hline
lightest neutralino & $m_{\tilde\chi^0_1}\ge35\mbox{ GeV}$
\\
\hline
\end{tabular}
\end{center}
\end{table}

Due to the weaker experimental bounds (in Table~\ref{tab1}) on sneutrino 
masses, 
the dominant contributions to $Z\rightarrow \ell_I \ell_J$ will come
from the diagrams mediated by chargino--sneutrino (see
Fig.~\ref{fig2}). Note that sneutrino masses can
be substantially lighter than charged slepton masses for large 
$\tan \beta$ and light SUSY--breaking masses, 
\bea
m^2_{\tilde\nu} &\approx& m^2_L + {1\over 2} M_Z^2 \cos 2\beta\;, \nn\\
m^2_{\tilde \ell_L} &\approx& m^2_L + \left( -{1\over 2} + s_W^2\right)
M_Z^2 \cos 2\beta \;,
\eea
which tends to increase the maximum relative contribution of 
chargino--sneutrino diagrams. 

We would like to emphasize that 
our results will depend on contributions with opposite signs that
often cancel when varying a parameter. For example, one would expect 
that the process $Z\rightarrow \ell_I \ell_J$ 
is optimized for light slepton masses. However, 
we observe frequently the opposite effect. Its branching ratio can 
increase by raising the mass of the sleptons up to values of 500 GeV, 
and only at masses above $1-2$ TeV the asymptotic regime 
is reached (see Fig.~\ref{fig2}). 
These cancellations give a one or two orders of magnitude uncertainty 
to any naive estimate, and underline the need for a complete 
calculation like the one presented here.

\begin{figure}
\centerline{\epsfig{file=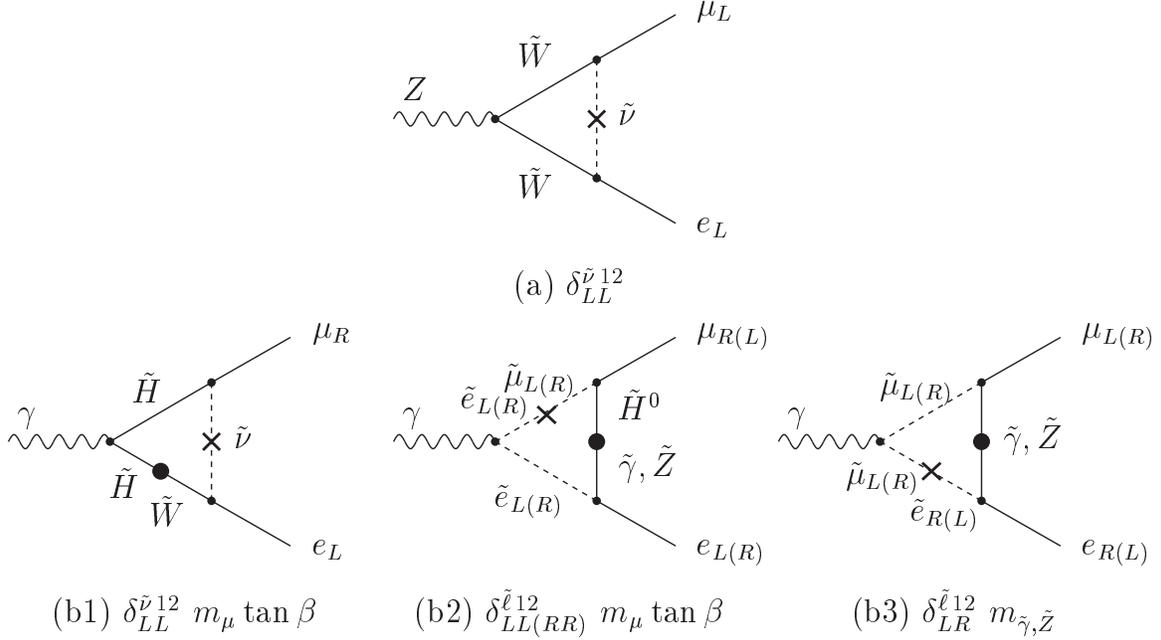,width=0.9\linewidth}}
\caption{Dominant diagrams contributing to (a) $Z\to\mu e$ and (b)
$\mu\to e\gamma$, in terms of gauginos, higgsinos and current eigenstates, 
showing the approximate linear dependence on the flavor--changing mass
insertions $\delta^{12}$ 
(crosses), the fermion mass insertions (big dots) and $\tan\beta$.
\label{fig3}}
\end{figure}

We give in Fig.~\ref{fig3} the dominant diagrams in terms of
gauginos, current eigenstates and mass insertions, specifying
the chirality of the external fermion. All the diagrams contributing
to $\ell_J\rightarrow\ell_I \gamma$ except for the last one grow with 
$\tan\beta$.


\section{Results}

\subsection{$Z\rightarrow \ell_I \ell_J$ at TESLA GigaZ}
 
Let us consider the process $Z\rightarrow \ell_I \ell_J$ uncorrelated
from other LFV processes. For SUSY masses
above the current limits it is possible to have 
$Z\rightarrow \mu e ; \tau e; \tau\mu$ 
at the reach of GigaZ.
The maximun rate is obtained when the 
second slepton $\tilde \ell_J$ is very heavy 
(i.e. $\delta^{IJ}\rightarrow \infty$). The largest contribution
comes from virtual sneutrino--chargino diagrams (all other contributions
are at least one order of magnitude smaller). It gives  
${\rm BR}(Z\rightarrow \ell_I \ell_J)$ from $2.5\times 10^{-8}$ for 
$\tan\beta=2$ to $7.5\times 10^{-8}$ for $\tan\beta=50$, practically 
independent of the lepton masses. The variation
is due to the mild dependence of chargino and
sneutrino masses on $\tan \beta$. These branching ratios are above
the values given in Eq.~(\ref{br4}). We find that a branching ratio larger than
$2\times 10^{-9}$ ($2\times 10^{-8}$) can be obtained with sneutrino
masses of up to 305 GeV (85 GeV) and chargino masses of up to
270 GeV (105 GeV).

Most of these values of ${\rm BR}(Z\rightarrow \ell_I \ell_J)$,
however, are correlated with an experimentally excluded rate of 
$\ell_J\rightarrow \ell_I \gamma$. We give below the results 
in the two scenarios (independent off-diagonal terms and maximal
mixing of the three flavors) described in the previous section.

{\it (i)} We separate the contribution of each $\delta^{IJ}$
setting all the other to zero. 
For the first two families, after 
scanning for all the parameters in the model we find that 
${\rm BR}(\mu\rightarrow e \gamma)<1.2\times 10^{-11}$ implies
${\rm BR}(Z\rightarrow \mu e)<1.5\times 10^{-10}$, which 
is below the reach of GigaZ.

\begin{figure}
\begin{center}
\begin{tabular}{cc}
\epsfig{file=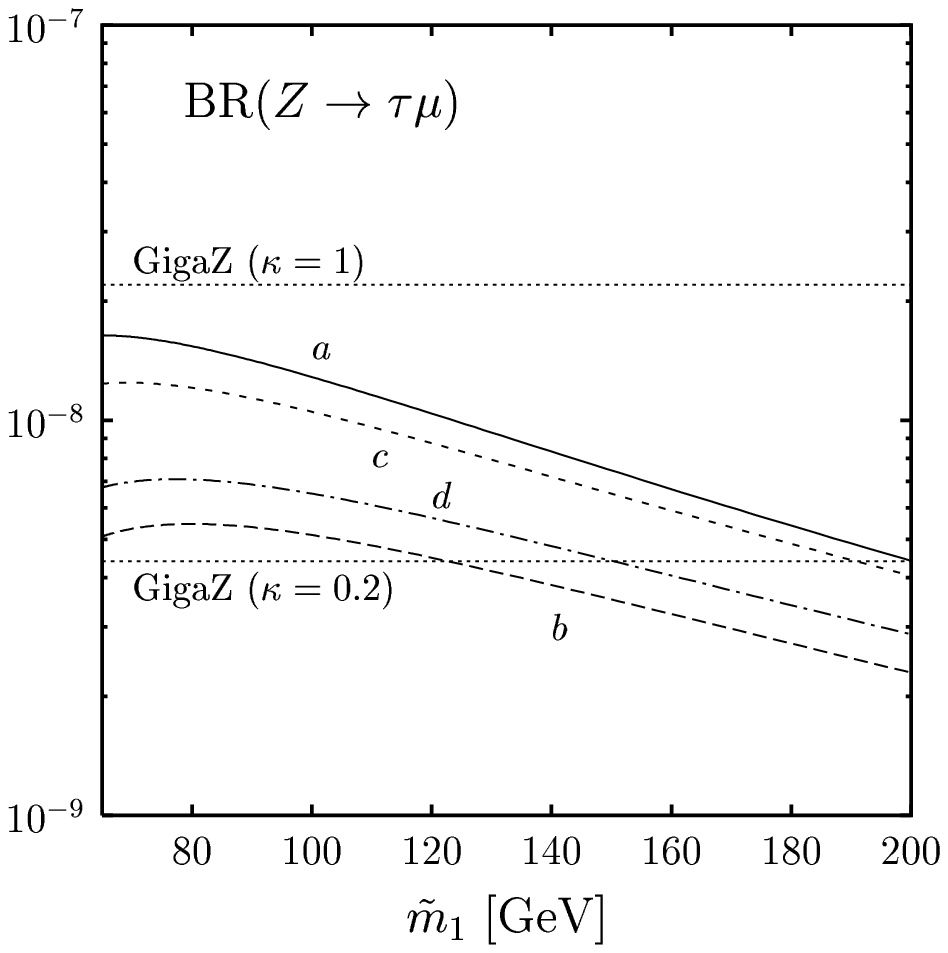,width=0.47\linewidth} &
\epsfig{file=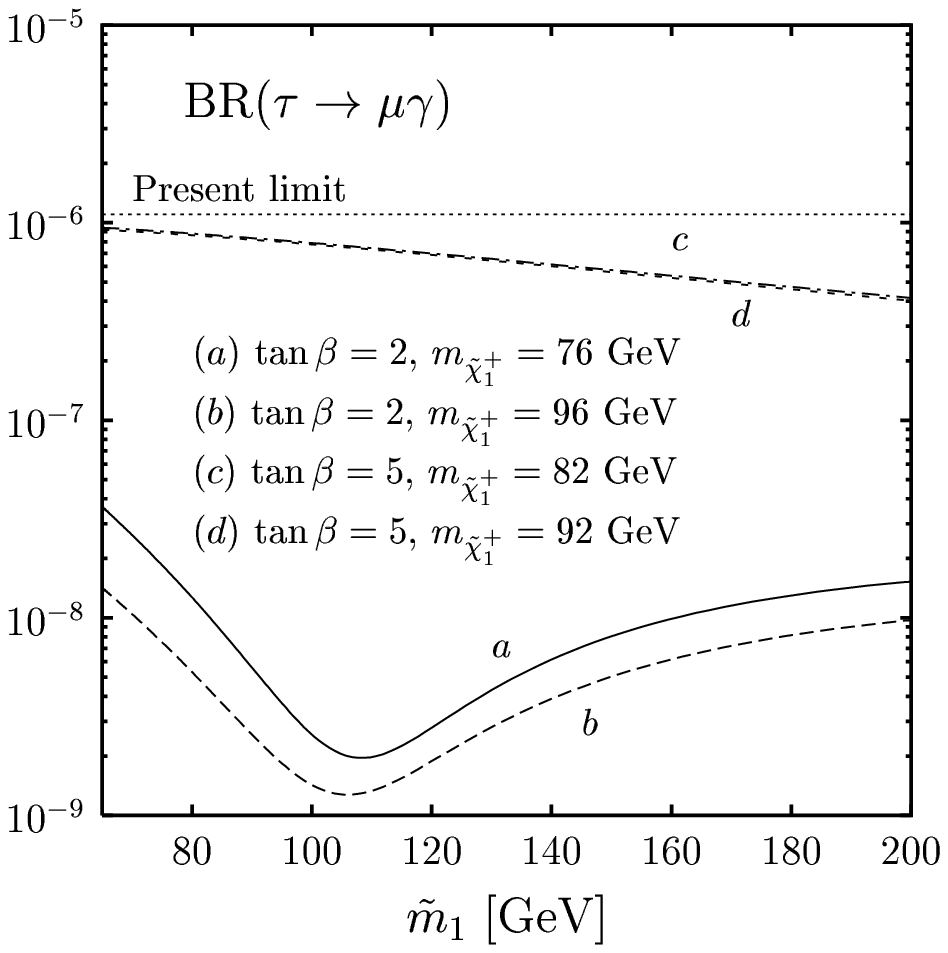,width=0.47\linewidth}
\end{tabular}
\end{center}
\caption{BR$(Z\to\tau\mu)$ and $BR(\tau\to\mu\gamma)$
 as a function of the lightest sneutrino mass ($\tilde m_1$) with the
other one decoupled ($\delta^{\tilde\nu\,23}_{LL}\to\infty$), in several 
SUSY scenarios at the reach of GigaZ.
\label{fig4}}
\end{figure}

A more promising result is obtained for the processes involving
the $\tau$ lepton. It turns out (see also next section) that 
the bounds from $\tau\rightarrow \ell_I \gamma$ can be
avoided while still keeping a rate of 
$Z\rightarrow \tau \ell_I$ at the reach of the best GigaZ
projection (see Fig.~\ref{fig4}). In particular, for large 
$\delta^{\tilde\nu\,13}_{LL}$ (or $\delta^{\tilde\nu\,23}_{LL}$) and a 
light sneutrino (of around 
70 GeV) we get ${\rm BR}(Z\rightarrow \tau e) \approx 
1.6\times 10^{-8}$
for ${\rm BR}(\tau\rightarrow e \gamma)\approx 
3.5\times 10^{-8}$, which is two orders of magnitude below current limits
(with similar results for ${\rm BR}(Z\rightarrow \tau\mu)$ and 
${\rm BR}(\tau\rightarrow \mu \gamma)$).
This result is due to the sneutrino--chargino diagram.
The contributions due to charged slepton mixing are essentially different 
in the sense that they saturate the experimental bound to 
$\tau\rightarrow \ell_I \gamma$
giving a small effect (at most, one order of magnitude below the
reach of GigaZ) in $Z\rightarrow \tau \ell_I$.
We obtain events at the reach of GigaZ with lightest sneutrino 
masses from 55 to 215 GeV, lightest chargino from 75 to 100 GeV, and 
$\tan\beta$ up to 7.

{\it (ii)} In the case with maximal mixing between the three slepton
flavors it is not consistent to take $\delta^{IJ}\not= 0$ 
and $\delta^{IK}=\delta^{KJ}=0$. In terms of slepton mass differences,
only two of the three mass differences are independent 
($\delta^{23}=\delta^{13}-\delta^{12}$).
In terms of off-diagonal terms $\tilde \delta^{IJ}$ in the mass 
matrix, for maximal mixing only one 
of them can be put to zero. Note that in this case the non-observation
of $\mu\rightarrow e \gamma$ will constraint all the $\delta^{IJ}$
parameters, not only $\delta^{12}$: a non-diagonal $\tilde \delta^{12}$  
mass insertion would be generated through a 
$\tilde \delta^{13}$ followed by a $\tilde \delta^{32}$.
In fact, we find that the constraints from 
$\tau\rightarrow e \gamma; \mu\gamma$
are always weaker than the one from $\mu\rightarrow e \gamma$.
A branching ratio ${\rm BR}(\mu\rightarrow e \gamma)<1.2\times 10^{-11}$ 
implies
${\rm BR}(Z\rightarrow \mu e;\;\tau e;\;\tau \mu)\lsim 10^{-9}$,
and the three lepton flavor violating decays of the $Z$ boson would
be out of the reach of Giga Z.


\subsection{Bounds on $\delta^{IJ}$ from $\ell_J\rightarrow \ell_I \gamma$}

The bounds on the $\delta^{IJ}$ parameters 
establish how severe is the flavor problem in the
lepton sector of the MSSM. We will update them here, including the 
sneutrino--chargino contributions neglected in previous works \cite{Masiero}
and the general slepton--neutralino contributions (photino diagrams are
typically subdominant as pointed out by Ref.~\cite{Feng:2001sq}).
In addition, we also consider the case of maximum mixing between the
three slepton families. 

\renewcommand{\arraystretch}{1}
\begin{table}[p]
\caption{Bounds on the $\delta^{12}$s from
BR$(\mu\to e\gamma) < 1.2\times 10^{-11}$ in different SUSY scenarios
assuming no mixing with the third family.
The bounds on $\delta^{IJ}$ in the case of three family mixing can be
read from these ones (see text).
\label{tab2}}
\begin{center}
\begin{tabular}{|c|c|c||c|c|c|c|}
\hline
\boldmath$\delta^{\tilde\nu\,12}_{LL}$ 
& $\tilde{m}_1$ & $M_2$ & $\mu=-500$ & $\mu=-150$ & $\mu=150$ & $\mu= 500$ \\
\hline\hline
   & 100 & 150 	& $14\times 10^{-3}$ 
		& $1.0\times 10^{-3}$
		& $0.3\times 10^{-3}$
		& $1.0\times 10^{-3}$ \\
\cline{3-7}
$\tan\beta=2$ 
   &     & 500  & $33\times 10^{-3}$ 
		& $3.0\times 10^{-3}$
		& $1.7\times 10^{-3}$
		& $13\times 10^{-3}$ \\
\cline{2-7}
   & 500 & 150  & $3.7\times 10^{-3}$ 
		& $1.1\times 10^{-3}$
		& $1.2\times 10^{-3}$
		& $1.7\times 10^{-3}$ \\
\cline{3-7}
   &     & 500  & $7.3\times 10^{-3}$ 
		& $2.6\times 10^{-3}$
		& $2.3\times 10^{-3}$
		& $4.1\times 10^{-3}$ \\
\hline
  & 100 & 150 	& $9.3\times 10^{-5}$ 
		& $2.1\times 10^{-5}$
		& $2.0\times 10^{-5}$
		& $8.6\times 10^{-5}$ \\
\cline{3-7}
$\tan\beta=50$ 
   &     & 500  & $80\times 10^{-5}$ 
		& $9.0\times 10^{-5}$
		& $8.8\times 10^{-5}$
		& $77\times 10^{-5}$ \\
\cline{2-7}
   & 500 & 150  & $9.7\times 10^{-5}$ 
		& $4.5\times 10^{-5}$
		& $4.5\times 10^{-5}$
		& $9.4\times 10^{-5}$ \\
\cline{3-7}
   &     & 500  & $22\times 10^{-5}$ 
		& $9.6\times 10^{-5}$
		& $9.5\times 10^{-5}$
		& $21\times 10^{-5}$ \\
\hline
\hline
\hline 
\boldmath$\delta^{\tilde\ell\,12}_{LL}$
& $\tilde{m}_1$ & $M_2$ & $\mu=-500$ & $\mu=-150$ & $\mu=150$ & $\mu= 500$ \\
\hline
  & 100 & 150 	& $7.5\times 10^{-3}$ 
		& $2.6\times 10^{-3}$
		& $1.7\times 10^{-3}$
		& $5.0\times 10^{-3}$ \\
\cline{3-7}
 $\tan\beta=2$  
   &     & 500  & $84\times 10^{-3}$ 
		& $11\times 10^{-3}$
		& $7.0\times 10^{-3}$
		& $41\times 10^{-3}$ \\
\cline{2-7}
   & 500 & 150  & $14\times 10^{-3}$ 
		& $8.5\times 10^{-3}$
		& \boldmath$0.12$
		& $32\times 10^{-3}$ \\
\cline{3-7}
   &     & 500  & $24\times 10^{-3}$ 
		& $20\times 10^{-3}$
		& $32\times 10^{-3}$
		& $26\times 10^{-3}$ \\
\hline
   & 100 & 150 	& $2.4\times 10^{-4}$ 
		& $0.8\times 10^{-4}$
		& $0.8\times 10^{-4}$
		& $2.4\times 10^{-4}$ \\
\cline{3-7}
$\tan\beta=50$ 
   &     & 500  & $24\times 10^{-4}$ 
		& $3.5\times 10^{-4}$
		& $3.4\times 10^{-4}$
		& $23\times 10^{-4}$ \\
\cline{2-7}
   & 500 & 150  & $7.4\times 10^{-4}$ 
		& $6.4\times 10^{-4}$
		& $6.9\times 10^{-4}$
		& $7.6\times 10^{-4}$ \\
\cline{3-7}
   &     & 500  & $9.7\times 10^{-4}$ 
		& $9.6\times 10^{-4}$
		& $9.8\times 10^{-4}$
		& $9.7\times 10^{-4}$ \\
\hline
\hline
\hline
\boldmath$\delta^{\tilde\ell\,12}_{RR}$
& $\tilde{m}_1$ & $M_2$ & $\mu=-500$ & $\mu=-150$ & $\mu=150$ & $\mu= 500$ \\
\hline\hline
   & 100 & 150 	& $4.2\times 10^{-3}$ 
		& $1.5\times 10^{-3}$
		& $1.8\times 10^{-3}$
		& $3.7\times 10^{-3}$ \\
\cline{3-7}
$\tan\beta=2$
   &     & 500  & $11\times 10^{-3}$ 
		& $3.2\times 10^{-3}$
		& $2.5\times 10^{-3}$
		& $8.3\times 10^{-3}$ \\
\cline{2-7}
   & 500 & 150  & $22\times 10^{-3}$ 
		& $10\times 10^{-3}$
		& $22\times 10^{-3}$
		& \boldmath$0.33$ \\
\cline{3-7}
   &     & 500  & $19\times 10^{-3}$
		& $12\times 10^{-3}$
		& \boldmath$0.33$ 
		& $35\times 10^{-3}$ \\
\hline
   & 100 & 150 	& $1.6\times 10^{-4}$ 
		& $0.6\times 10^{-4}$
		& $0.6\times 10^{-4}$
		& $1.5\times 10^{-4}$ \\
\cline{3-7}
$\tan\beta=50$
   &     & 500  & $3.8\times 10^{-4}$ 
		& $1.1\times 10^{-4}$
		& $1.1\times 10^{-4}$
		& $3.8\times 10^{-4}$ \\
\cline{2-7}
   & 500 & 150  & $16\times 10^{-4}$ 
		& $13\times 10^{-4}$
		& $15\times 10^{-4}$
		& $17\times 10^{-4}$ \\
\cline{3-7}
   &     & 500  & $9.4\times 10^{-4}$ 
		& $8.1\times 10^{-4}$
		& $8.7\times 10^{-4}$
		& $9.6\times 10^{-4}$ \\
\hline
\hline
\hline 
\boldmath$\delta^{\tilde\ell\,12}_{LR}$
& $\tilde{m}_1$ & $M_2$ & $\mu=-500$ & $\mu=-150$ & $\mu=150$ & $\mu= 500$ \\
\hline\hline
   & 100 & 150 	& $1.6\times 10^{-6}$ 
		& $1.5\times 10^{-6}$
		& $1.6\times 10^{-6}$
		& $1.7\times 10^{-6}$ \\
\cline{3-7}
$\tan\beta=2$
   &     & 500  & $4.5\times 10^{-6}$ 
		& $4.4\times 10^{-6}$
		& $4.7\times 10^{-6}$
		& $4.6\times 10^{-6}$ \\
\cline{2-7}
   & 500 & 150  & $1.3\times 10^{-6}$ 
		& $1.2\times 10^{-6}$
		& $1.2\times 10^{-6}$
		& $1.2\times 10^{-6}$ \\
\cline{2-7}
   &     & 500  & $7.6\times 10^{-6}$ 
		& $7.5\times 10^{-6}$
		& $7.6\times 10^{-6}$
		& $7.7\times 10^{-6}$ \\
\hline
   & 100 & 150 	& $1.6\times 10^{-6}$ 
		& $1.5\times 10^{-6}$
		& $1.6\times 10^{-6}$
		& $1.6\times 10^{-6}$ \\
\cline{3-7}
$\tan\beta=50$
   &     & 500  & $4.5\times 10^{-6}$ 
		& $4.5\times 10^{-6}$
		& $4.6\times 10^{-6}$
		& $4.5\times 10^{-6}$ \\
\cline{2-7}
   & 500 & 150  & $1.3\times 10^{-6}$ 
		& $1.2\times 10^{-6}$
		& $1.2\times 10^{-6}$
		& $1.3\times 10^{-6}$ \\
\cline{3-7}
   &     & 500  & $7.7\times 10^{-6}$ 
		& $7.6\times 10^{-6}$
		& $7.6\times 10^{-6}$
		& $7.7\times 10^{-6}$ \\
\hline
\end{tabular}
\end{center}
\end{table}
\renewcommand{\arraystretch}{1.3}

The limits come exclusively from the process 
$\ell_J\rightarrow \ell_I \gamma$.
To estimate the MSSM prediction we combine low and 
high values of the relevant parameters: $\tan\beta=2;50$,
$\tilde m_1=100;500$ GeV, and the gaugino and higgsino mass 
parameters $M_2=150;500$ GeV and $\mu=\pm 150;\pm 500$ GeV.

{\it (i)} The results for the case with a decoupled family
are summarized in Table~\ref{tab2}.
We include the bounds from $\mu\rightarrow e
\gamma$ to $\delta^{\tilde \nu\,12}_{LL}$, 
$\delta^{\tilde \ell\,12}_{LL}$, $\delta^{\tilde \ell\,12}_{RR}$ 
and $\delta^{\tilde \ell\,12}_{LR}$. A $\delta \approx 10^{-3}$ 
implies a 1\permil\ degeneracy between the two slepton masses. We 
observe that the degeneracy between the selectron and the smuon
is required even for large SUSY masses, and it must be stronger
if $\tan\beta$ is large,
as expected from the diagrams in Fig.~\ref{fig3}. 
The small values of $\delta^{\tilde\ell\,12}_{LR}$, around $10^{-6}$, imply 
just that the scalar
trilinears, usually assumed proportional to the Yukawa couplings, are 
 small. Particularly weak bounds on the $\delta$'s (bold faced in 
Table~\ref{tab2}) are obtained when approaching a dip of the curves in 
Fig.~\ref{fig2}. 
This occurs for certain values of the SUSY parameters due to cancellations 
of the contributions of the various particles running in the loops.

The experimental bounds on the mass differences involving the 
third family are much weaker. They come from 
$\tau\rightarrow \ell_I \gamma$ (and not from $\mu\rightarrow e \gamma$),
since in this case we are asuming that the mixing with $\ell_J$ 
($J\not= I,3$) is negligible.
For small $\tan\beta$
we find no bounds on any $\delta^{I3}$ (except for 
$\delta^{\tilde\ell\,I3}_{LR}$). For large $\tan\beta$ the
bounds are (depending on the values of the SUSY--breaking masses)
$\delta^{\tilde \nu\,I3}_{LL} = 0.03 \;{\rm to}\; 1.3$; 
$\delta^{\tilde \ell\,I3}_{LL} = 0.14 \;{\rm to}\; \infty$; 
and 
$\delta^{\tilde \ell\,I3}_{RR} = 0.11 \;{\rm to}\; \infty$.
For the $LR$ mass insertions, we find
$\delta^{\tilde \ell\,I3}_{LR} = 0.05 \;{\rm to}\; \infty$,
independent of $\tan\beta$.
This results improve the bounds obtained in Ref.~\cite{Masiero}, 
in particular, the ones involving the second 
and third lepton families.

{\it (ii)} As explained before, if the three slepton families are 
maximally mixed (as 
suggested by the experiments on neutrino oscilations) 
then the strongest bounds on all the slepton mass differences come
from $\mu\rightarrow e \gamma$ exclusively. As we will see, however,
the bounds can be read directly from Table~\ref{tab2}, since
the small values obtained for $\delta^{12}$ admit an analysis 
based on mass insertions. Let us first suppose that the first
and second sneutrino families are degenerated 
($\delta^{\tilde \nu\,12}_{LL}=0$).
Then a mass difference 
$\delta^{\tilde \nu\,13}_{LL}=\delta^{\tilde \nu\,23}_{LL}=\delta$ 
generates an off-diagonal mass term (see Appendix B3) 
$\tilde \delta^{\tilde \nu\,12}_{LL}=\delta/4$. In consecuence, 
the degree of 
mass degeneracy with the third family ($\delta$)
imposed by $\mu\rightarrow e \gamma$ is just four times the value
given in Table~\ref{tab2}. We have checked numerically that this estimate
is quite accurate. Analogously, the bounds on 
$\delta^{\tilde \nu\,12}_{LL}=\delta^{\tilde \nu\,13}_{LL}$ when 
$\delta^{\tilde \nu\,23}_{LL}=0$ and 
on $\delta^{\tilde \nu\,12}_{LL}=\delta^{\tilde \nu\,23}_{LL}$ when 
$\delta^{\tilde \nu\,13}_{LL}=0$ are
respectively two and four times the values in Table~\ref{tab2}.
In the same way we can read there the bounds on 
$\delta^{\tilde \ell\,IJ}_{LL}$ and 
$\delta^{\tilde \ell\,IJ}_{RR}$, which establish the degree
of mass degeneracy between the three families of charged sleptons.

\subsection{Lepton flavor violation and $g_\mu-2$}

Finally we would like to comment on the relation between
$\mu\rightarrow e \gamma$ and the muon anomalous magnetic dipole 
moment. 
See Ref.~\cite{Carvalho:2001ex} for more exhaustive analyses of the
constraints on lepton flavor violation in the MSSM from the muon  
anomalous magnetic moment measurement.
A $g_\mu-2$ correction would be generated by the diagrams
in Fig.~\ref{fig3}b if no mass insertions $\delta^{\tilde \nu\,IJ}_{LL}$,
$\delta^{\tilde \ell\,IJ}_{LL}$, $\delta^{\tilde \ell\,IJ}_{RR}$ are 
included and $\delta^{\tilde \ell\,IJ}_{LR}$ is replaced by 
$\delta^{\tilde \ell\,22}_{LR}$.
In this sense, $g_\mu-2$ is a normalization of the
branching ratio BR$(\ell_J\rightarrow \ell_I \gamma)$ for processes changing
the muon flavor.

\begin{figure}
\begin{center}
\epsfig{file=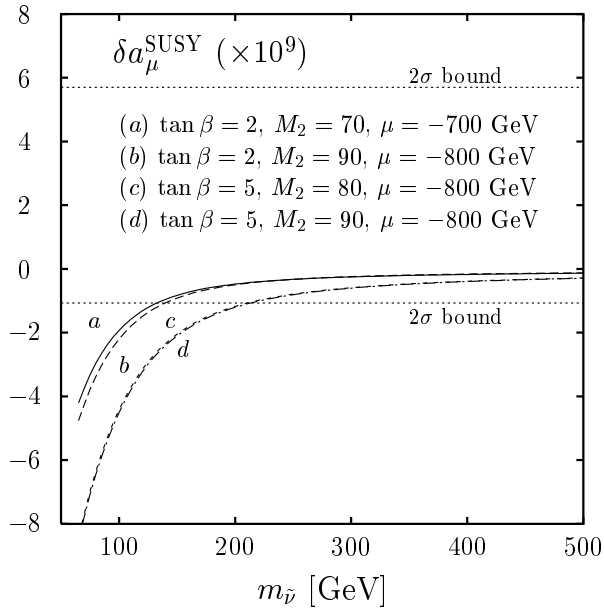,width=0.47\linewidth}
\end{center}
\caption{Total SUSY contribution to the muon anomalous magnetic dipole moment 
as a function of the sneutrino mass with
$m_L=m_R$, in the same SUSY scenarios as in Fig.~\ref{fig4}.
\label{fig5}}
\end{figure}

We plot in Fig.~\ref{fig5} the value of $a_\mu=(g_\mu-2)/2$ for the SUSY 
parameters in
the region accessible to GigaZ not excluded by $\tau\rightarrow
\mu\gamma$, taking for simplicity equal soft--breaking terms $m_L=m_R$ 
(they would not very different, for example, 
assuming left--right unification at the GUT scale). 
We obtain, in agreement with \cite{Moroi:1996yh}, positive or negative 
contributions correlated with the sign of the Higgsino mass parameter 
$\mu$ and similar in size to the weak corrections.  
The recently revised SM prediction \cite{Knecht:2001qf},
$a^{\rm SM}_\mu=11\ 659\ 179.2\ (9.4)\times 10^{-10}$, 
compared to the world average after the last data from the Brookhaven 
E821 experiment \cite{Brown:2001mg},
$a^{\rm exp}_\mu=11\ 659\ 202.3\ (15.1)\times 10^{-10}$, 
exhibits a 1.4$\sigma$ discrepancy:
$\delta a_\mu=a^{\rm exp}_\mu-a^{\rm SM}_\mu=(23.1\pm16.9)\times 10^{-10}$.
This indicates that the muon dipole moment may still need non--standard
contributions of positive sign. In any case, 
the MSSM contribution $\delta a^{\rm SUSY}_\mu$ is bounded at two standard
deviations by the dotted 
lines in Fig.~\ref{fig5}. Only the 
regions with heavier masses in the scenarios of Fig.~\ref{fig4} are favored. 

\begin{figure}
\begin{center}
\begin{tabular}{cc}
\epsfig{file=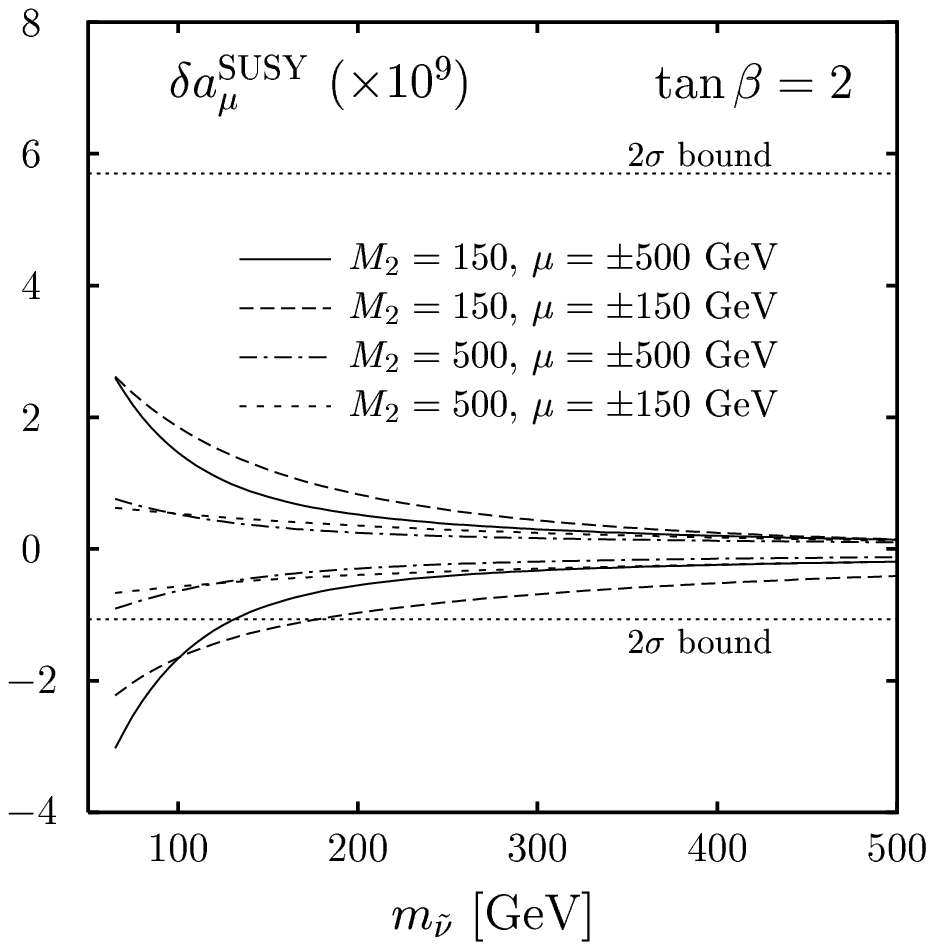,width=0.47\linewidth} &
\epsfig{file=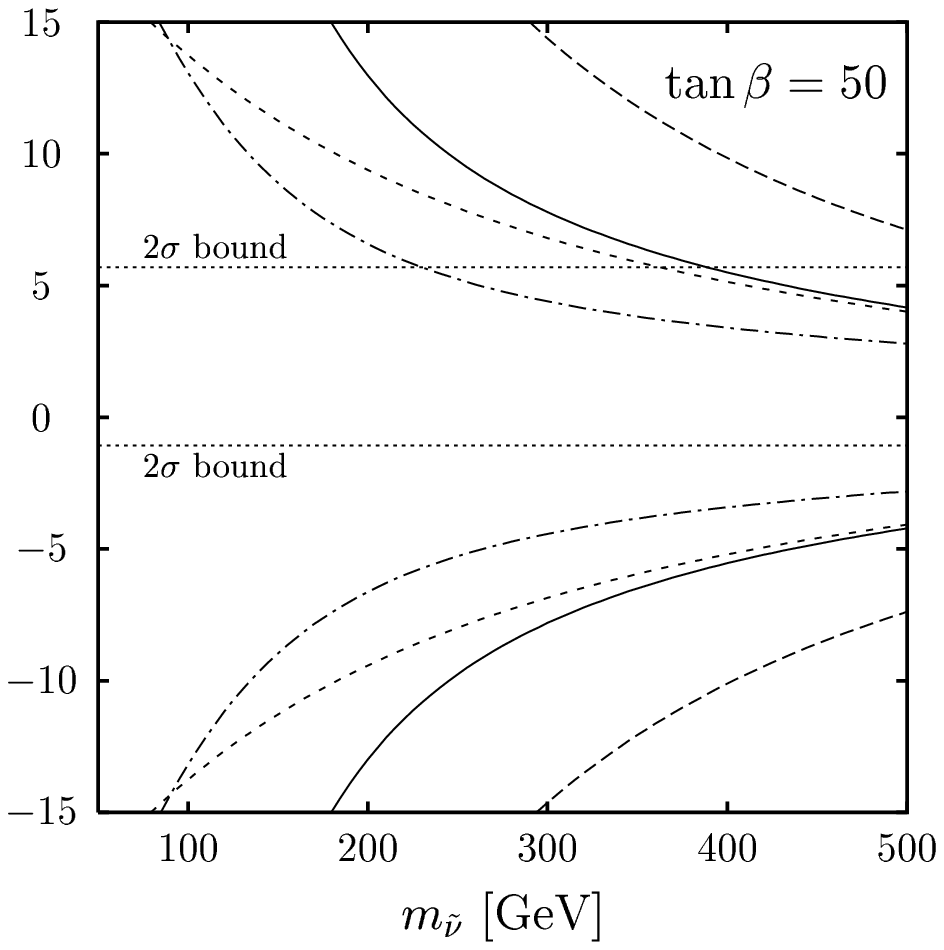,width=0.47\linewidth}
\end{tabular}
\end{center}
\caption{Total SUSY contribution to the muon anomalous magnetic dipole moment 
as a function of the sneutrino mass, with $m_L=m_R$.
\label{fig6}}
\end{figure}

Fig.~\ref{fig6} shows the muon anomalous magnetic moment for the different 
sets of SUSY parameters employed to explore the muon LFV decay in 
Table~\ref{tab2}.
Low values of $\tan\beta$ and positive values of $\mu$ are 
preferred by $g_\mu-2$, which implies less stringent bounds on the 
$\delta$ insertions parametrizing the flavor--changing lepton decays.


\section{Conclusions}

SUSY models introduce LFV corrections which are proportional to 
slepton mass squared differences. We have shown that the 
non--observation of $\mu\rightarrow e \gamma$ implies around a
1\permil\ degeneracy between the masses of the sleptons in the first
two families. Once this degeneracy is imposed, the
rate of $Z\rightarrow \mu e$ is always below the limits to be explored
at GigaZ. Moreover, if the mixing between the three slepton 
families is large then also the third family must
be (at the 1\permil) degenerated, and the processes
$Z\rightarrow \tau e;\;\tau\mu$ will not be observed at the GigaZ.
The degeneracy between the lightest
slepton families could be justified by the weakness of its Yukawa 
couplings, but for the third family it should put constraints 
on definite SUSY models. 

In contrast, if the mixing between the third and the first
slepton families is small, then the third family could be
much heavier than the other two and there would be no flavor
problem in the slepton sector (the bounds would come only
from $\tau\rightarrow \mu\gamma$, not from 
$\mu\rightarrow e \gamma$).
In this case, if the GigaZ option of TESLA
reaches its best projected sensitivity it could observe
$Z\rightarrow \tau\mu$ coming from the virtual exchange
of wino--sneutrino.

\vspace{2mm}
\noindent
{\bf Acknowledgments}\\
JII acknowledges the Theory Group of DESY Zeuthen and MM the Institute
for Nuclear Research (Moscow) for their kind hospitality. 
We thank F. del Aguila, A. Bueno, D. Gorbunov, J. Prades and T. Riemann 
for useful discussions.
This work has been supported by CICYT, Junta de Andaluc{\'\i}a and the 
European Union under contracts FPA2000-1558, FQM-101 and HPRN-CT-2000-00149,
respectively.

\noindent
{\bf Note added}\\
After the completion of this paper,
new data from BNL $g-2$ experiment \cite{Bennett:2002jb} appeared confirming 
the previous measurement with twice the precision. The discrepancy with 
the SM prediction is now more significant, up to $2.6\sigma$. 
There are also new experimental bounds on $\tau\to\mu\gamma$ 
\cite{Inami:2002us}. The new data do not introduce any qualitative changes
in our results.

\appendix

\section{Generic expressions at one loop for $Z\bar \ell_I \ell_J$}

\subsection{Feynman rules in terms of generic vertex couplings}
 
Let $f$ be a fermion, $\phi$ a scalar field and $P_{R,L}=(1\pm\gamma_5)/2$. 
The Feynman rules for the three vertex topologies needed are:

\begin{tabular}{lcl}
$\bullet$
Vertex $Z\bar f_Af_B$ &:&
$ig\gamma^\mu(g^L_{AB}P_L + g^R_{AB}P_R)\; ,$
\\
$\bullet$
Vertex $Z\phi_X^\dagger(p_2)\phi_Y(p_1)$ &:&
$igG_{XY}(p_1+p_2)^\mu\; ,$
\\
$\bullet$
Vertex $\phi_X^\dagger\bar f_Af_I$ &:&
$ig(c^L_{IAX}P_L + c^R_{IAX}P_R)\; .$
\end{tabular}

\subsection{Invariant amplitude}

The most general invariant amplitude for on--shell external legs is
\bea
{\cal M}=
-ig\frac{\alpha_W}{4\pi} \varepsilon^\mu_Z \bar u_{\ell_I}(p_2) 
\left[\gamma_\mu(f_V-f_A\gamma_5)
     +\frac{\sigma_{\mu\nu}q^\nu}{M_W}(if_M+f_E\gamma_5)
      \right] u_{\ell_J}(p_1)\; .
\eea

\noindent
Let us introduce the squared mass ratios $\lambda_n=m^2_n/M^2_W$ and
the dimensionless two-- and three--point one--loop integrals
\bea
\b_1(\lambda_0,\lambda_1)&\equiv& B_1(0;m^2_0,m^2_1)\; ,\\
\c_{..}(\lambda_0,\lambda_1,\lambda_2)&\equiv&
M^2_W\ C_{..}(0,M^2_Z,0;m^2_0,m^2_1,m^2_2)\; ,
\eea
from the usual tensor integrals \cite{'tHooft:1972fi,Passarino:1979jh},
\bea
B^\mu(p^2;m^2_0,m^2_1)&=&p^\mu B_1\; , 
\\
C^\mu(p^2_1,q^2,p^2_2;m^2_0,m^2_1,m^2_2)&=&p^\mu_1 C_{11}\! +\! p^\mu_2 C_{12}
\; ,
\\
C^{\mu\nu}(p^2_1,q^2,p^2_2;m^2_0,m^2_1,m^2_2)
&=&p^\mu_1 p^\nu_1 C_{21} + p^\mu_2 p^\nu_2 C_{22} 
+ (p^\mu_1 p^\nu_2 + p^\mu_2 p^\nu_1) C_{23}
+ g^{\mu\nu} M^2_W C_{24}\; . 
\eea
Note that $\c_0$, $\c_{23}$, $\c_{24}$, $\c_{11}+\c_{12}$ and $\c_{21}+\c_{22}$
are symmetric under the replacements $\lambda_1\leftrightarrow\lambda_2$, while
$\c_{11}-\c_{12}$ and $\c_{21}-\c_{22}$ are antisymmetric.
The form factors for each type of diagrams (Fig.~\ref{fig1}) are:

$\bullet$ 
Diagram of type A:
\bea
f_L^{\chi\chi s}&=& \sum_{XAB} \Bigg\{ g^L_{AB} c^{L^*}_{IAX} c^L_{JBX}\
	      \sqrt{\lambda_A \lambda_B}\
	      \c_0(\lambda_X,\lambda_A,\lambda_B)
\nn\\
          &&+ g^R_{AB} c^{L^*}_{IAX} c^L_{JBX}\ \Bigg[
	      \lambda_Z\ \c_{23}(\lambda_X,\lambda_A,\lambda_B) 
	      -2\ \c_{24}(\lambda_X,\lambda_A,\lambda_B)
	      +\frac{1}{2} \Bigg] \Bigg\}\; ,
\\
f_R^{\chi\chi s}&=& f_L^{\chi\chi s}\ (\ L \leftrightarrow R\ )\; ,       
\\
f_M^{\chi\chi s}
	&=& \sum_{XAB}\Bigg\{ \frac{\sqrt{\lambda_J}}{2}\
	     [ (g^R_{AB} c^{L^*}_{IAX} c^L_{JBX}
	       + g^L_{AB} c^{R^*}_{IAX} c^R_{JBX}) 
	       + (I\leftrightarrow J)^* ]\nn\\ 
	&&\hspace{18mm}\times[
	       	\c_{11}(\lambda_X,\lambda_A,\lambda_B)+
		\c_{21}(\lambda_X,\lambda_A,\lambda_B)+
		\c_{23}(\lambda_X,\lambda_A,\lambda_B)]\nn\\
	&&\hspace{8mm}+\frac{\sqrt{\lambda_A}}{2}\ 
              [ (g^L_{AB} c^{L^*}_{IAX} c^R_{JBX}
	       + g^R_{AB} c^{R^*}_{IAX} c^L_{JBX}) 
	       + (I\leftrightarrow J)^* ]\ 
	      \c_{12}(\lambda_X,\lambda_A,\lambda_B)\Bigg\}\; ,
\\
if_E^{\chi\chi s}
	&=& \sum_{XAB}\Bigg\{ \frac{\sqrt{\lambda_J}}{2}\
	     [ (g^R_{AB} c^{L^*}_{IAX} c^L_{JBX}
	       - g^L_{AB} c^{R^*}_{IAX} c^R_{JBX}) 
	       - (I\leftrightarrow J)^* ]\nn\\ 
	&&\hspace{18mm}\times[
	       	\c_{11}(\lambda_X,\lambda_A,\lambda_B)+
		\c_{21}(\lambda_X,\lambda_A,\lambda_B)+
		\c_{23}(\lambda_X,\lambda_A,\lambda_B)]\nn\\
	&&\hspace{8mm}+\frac{\sqrt{\lambda_A}}{2}\ 
              [  (g^L_{AB} c^{L^*}_{IAX} c^R_{JBX}
	       - g^R_{AB} c^{R^*}_{IAX} c^L_{JBX})
               - (I\leftrightarrow J)^* ]\ 
	      \c_{12}(\lambda_X,\lambda_A,\lambda_B)\Bigg\}\; .
\eea

$\bullet$
Diagram of type B:
\bea
f_L^{ss\chi}&=&       -2 \sum_{AXY} G_{XY} c^{L^*}_{IAX} c^L_{JAY}\ 
             \c_{24}(\lambda_A,\lambda_X,\lambda_Y)\; , 
\\
f_R^{ss\chi}&=&        f_L^{ss\chi}\ (\ L \leftrightarrow R\ )\; ,    
\\
f_M^{ss\chi}&=&  \sum_{AXY}\Bigg\{ -\frac{\sqrt{\lambda_J}}{2}\
	    [ G_{XY}(c^{L^*}_{IAX} c^L_{JAY}+ c^{R^*}_{IAX} c^R_{JAY})
		+(I\leftrightarrow J)^* ]\nn\\
	    &&\hspace{23mm}\times [\c_{11}(\lambda_A,\lambda_X,\lambda_Y)+
          	      \c_{12}(\lambda_A,\lambda_X,\lambda_Y)+
	  	      \c_{23}(\lambda_A,\lambda_X,\lambda_Y)] \nn\\
	    &&\hspace{13mm}+\frac{\sqrt{\lambda_A}}{2}\ 
	    [ G_{XY} c^{L^*}_{IAX} c^R_{JAY} + (I\leftrightarrow J)^*]
\nn\\   
&&\hspace{23mm}\times [\c_0(\lambda_A,\lambda_X,\lambda_Y)+
          \c_{11}(\lambda_A,\lambda_X,\lambda_Y)+
	  \c_{12}(\lambda_A,\lambda_X,\lambda_Y)]\Bigg\}\; ,
\\
if_E^{ss\chi}&=&  \sum_{AXY}\Bigg\{ -\frac{\sqrt{\lambda_J}}{2}\
	    [ G_{XY} (c^{L^*}_{IAX} c^L_{JAY}- c^{R^*}_{IAX} c^R_{JAY})
		-(I\leftrightarrow J)^* ]\nn\\
	    &&\hspace{23mm}\times [\c_{11}(\lambda_A,\lambda_X,\lambda_Y)+
          	      \c_{12}(\lambda_A,\lambda_X,\lambda_Y)+
	  	      \c_{23}(\lambda_A,\lambda_X,\lambda_Y)] \nn\\
	    &&\hspace{13mm}+\frac{\sqrt{\lambda_A}}{2}\ 
	    [ G_{XY} c^{L^*}_{IAX} c^R_{JAY} - (I\leftrightarrow J)^*]
\nn\\   
&&\hspace{23mm}\times [\c_0(\lambda_A,\lambda_X,\lambda_Y)+
          \c_{11}(\lambda_A,\lambda_X,\lambda_Y)+
	  \c_{12}(\lambda_A,\lambda_X,\lambda_Y)]\Bigg\}\; .
\eea

$\bullet$
Diagram of type C:
\bea
f_L^{\chi s}&=&    -\frac{\cos2\theta_W}{2c_W} \sum_{AX} c^{L^*}_{IAX} 
c^L_{JAX}\
            \b_1(\lambda_A,\lambda_X)\; ,
\\
f_R^{\chi s}&=&      \frac{s^2_W}{c_W} \sum_{AX} c^{R^*}_{IAX} c^R_{JAX}\
            \b_1(\lambda_A,\lambda_X)\; ,
\\	  
f_M^{\chi s}&=&0\; ,
\\
f_E^{\chi s}&=&0\; .
\eea
The tensor integrals are numerically evaluated  with the computer
program {\tt LoopTools} \cite{Hahn:1999yk}, based on {\tt FF}
\cite{vanOldenborgh:1990wn}.

Non--trivial checks of our expressions are the finiteness of the
amplitude and the test of the decoupling of heavy particles running 
in the loops, that must take place both in the SM and the MSSM 
\cite{decoupling}. These conditions are fulfilled only
when summing over the different type of diagrams involved thanks to the
relations existing among vertex couplings. 
Note that the ultraviolet--divergent tensor integrals are the same
that diverge with a large mass $M$,
\bea
\c_{24}\to-\frac{1}{2\epsilon}-\frac{1}{2}\log M,\quad
\b_1\to\frac{1}{\epsilon}+\log M,\quad
\epsilon=D-4\; .
\eea
All the other tensor integrals are finite and vanish for large masses.

\section{Masses, mixings and vertex couplings in the MSSM}

Notation: the indices $I$ or $J$ refer to the flavor of the external fermion;
the indices $A$ or $B$ refer to a chargino/neutralino mass eigenstate
($\tilde\chi^\pm_{A=1,2}$ and $\tilde\chi^0_{A=1,2,3,4}$);
the indices $X$ or $Y$ refer to a charged slepton/sneutrino mass 
eigenstate ($\tilde\ell_{X=1,\dots,6}$ and $\tilde\nu_{X=1,2,3}$).

\subsection{Charged sleptons}

Let $\tilde\ell_{L_I}$ and $\tilde\ell_{R_I}$ be the superpartners of the
charged leptons $\ell_{L_I}$ and $\ell_{R_I}$, respectively.
The $6\times6$ mass matrix of three generations of (charged) sleptons can be 
written as
\bea
{\bf M}^2_{\tilde\ell}=\left(\ba{cc}{\bf m}^2_{LL} & {{\bf m}^{2}_{LR}}^T \\
			            {\bf m}^2_{LR} & {\bf m}^2_{RR} \ea\right) 
\; ,
\eea
where ${\bf m}^2_{LL}$ and ${\bf m}^2_{RR}$ are 
$3\times3$ hermitian matrices and ${\bf m}^{2}_{LR}$ is a $3\times3$ matrix, 
given by
\bea
({\bf m}^2_{LL})_{IJ}&=&({\bf m}^2_{L})_{IJ}+\left[m^2_{\ell_I}+
\left(-\frac{1}{2}+s^2_W\right)M^2_Z\cos2\beta\right]\delta_{IJ}\; ,
\\
({\bf m}^2_{RR})_{IJ}&=&({\bf m}^2_{R})_{IJ}+
		[m^2_{\ell_I}-M^2_Z\cos2\beta s^2_W]\delta_{IJ}\; ,
\\
({\bf m}^2_{LR})_{IJ}&=&({\bf A}_\ell)_{IJ}\frac{v\cos\beta}{\sqrt{2}}
                      -m_{\ell_I}\mu\tan\beta\ \delta_{IJ}\; .
\eea
The mass matrix 
${\bf M}^2_{\tilde\ell}$ can be diagonalized by a $6\times6$ unitary matrix 
${\bf S}^{\tilde\ell}$,
\bea
{\bf S}^{\tilde\ell}\ {\bf M}^2_{\tilde\ell}\ {{\bf S}^{\tilde\ell}}^\dagger=
{\bf diag}(m^2_{\tilde\ell_X}),\quad X=1,\dots,6\; .
\eea
The mass eigenstates are then given by
\bea
\tilde\ell_X={\bf S}^{\tilde\ell}_{X,I}  \tilde\ell_{L_I}
            +{\bf S}^{\tilde\ell}_{X,I+3}\tilde\ell_{R_I}\; ,
          \quad X=1,\dots,6,\ I=1,2,3\; .
\eea

\subsection{Sneutrinos}

There are only `left--handed' sneutrinos in the MSSM. Let $\tilde\nu_{L_I}$ be
the superpartner of the left-handed neutrino $\nu_I$. Then the $3\times3$
sneutrino mass matrix contains the same soft SUSY--breaking mass term
as the `left--handed' sleptons and a different $D$ term:
\bea
({\bf M}_{\tilde\nu}^2)_{IJ}=({\bf m}_{L}^2)_{IJ}+\frac{1}{2}M^2_Z
\cos2\beta\ \delta_{IJ}\; ,
\eea
and it is diagonalized by a $3\times3$ unitary matrix ${\bf S}^{\tilde\nu}$,
\bea
{\bf S}^{\tilde\nu}\ {\bf M}^2_{\tilde\nu}\ {{\bf S}^{\tilde\nu}}^\dagger=
{\bf diag}(m^2_{\tilde\nu_X}),\quad X=1,2,3\; ,
\eea
so that the sneutrino mass eigenstates are
\bea
\tilde\nu_X={\bf S}^{\tilde\ell}_{X,I}  \tilde\nu_{L_I},
          \quad X=1,2,3,\ I=1,2,3\; .
\eea

\subsection{Slepton matrices in terms of $\delta$ mass insertions}

{\it (i)}
Assuming that only two generations ($I$ and $J$) of charged sleptons 
mix and they do it maximally ($\theta=\pi/4$), only the following 
$4\times4$ symmetric mass matrix, with entries $I,J,I+3,J+3$, is 
relevant:
\bea
{\bf M}^2_{\tilde\ell}=\tilde m^2\left(\ba{cccc} 1 & \cdot & \cdot & \cdot \\
\delta^{\tilde\ell\;IJ}_{LL} & 1 & \cdot & \cdot \\
\delta^{\tilde\ell\;II}_{LR} & \delta^{\tilde\ell\;IJ}_{LR} & 1 & \cdot \\
\delta^{\tilde\ell\;JI}_{LR} & \delta^{\tilde\ell\;JJ}_{LR} & 
\delta^{\tilde\ell\;IJ}_{RR} & 1
\ea\right)\; .
\eea
The insertions $\delta^{\tilde\ell\;II}_{LR}$ and 
$\delta^{\tilde\ell\;JJ}_{LR}$ are flavor conserving. 
We assume that, alternatively, only one of these $\delta$'s is different 
from zero. Then, the relevant non-diagonal $2\times2$ submatrix:
\bea
{\bf m}^2=\tilde m^2\left(\ba{cc} 1 &\delta \\ \delta & 1 \ea\right)
\eea
is trivially diagonalized by the following submatrix of ${\bf S}$:
\bea
{\bf U}=\frac{1}{\sqrt{2}}\left(\ba{cc} 1 & -1 \\
			                1 &  1 \ea\right)\ ,
\eea
yielding the eigenvalues: 
\bea
\tilde m^2_{1,2}=\tilde m^2(\sqrt{1+\delta^2}\mp\delta)\; ,
\eea
where $\delta=(\tilde m^2_2-\tilde m^2_1)/(2\tilde m^2)$ is the mass 
splitting between both generations of sleptons.

The relevant $2\times2$ submatrix for the sneutrinos in terms of the mass 
insertion $\delta^{\tilde\nu\;IJ}_{LL}$ is constructed in a similar way.

{\it (ii)}
For the case when the three generations of sleptons mix we employ the
{\em standard parametrization} for the relevant $3\times3$ submatrix: one CP 
phase (that we set to zero) and three mixing angles $\theta_{12}$, 
$\theta_{13}$, $\theta_{23}$ where $\theta_{IJ}$ 
represents the mixing between familiess $I$ and $J$ when the mixing to the
remaining one is zero. We take again maximal mixing, $\theta_{12}=\theta_{13}=
\theta_{23}=\pi/4$. Then,
\bea
{\bf U}=\frac{1}{2\sqrt{2}}\left(\ba{ccc} \sqrt{2} & -\sqrt{2} & -2 \\
			\sqrt{2}-1 & \sqrt{2}+1 & -\sqrt{2} \\
			\sqrt{2}+1 & \sqrt{2}-1 & \sqrt{2} \ea\right)
\eea
is the unitary matrix that diagonalizes the symmetric mass matrix:
\bea
{\bf m}^2=m^2_1\left(\ba{ccc} \tilde\delta^{11} & \cdot & \cdot \\
			      \tilde\delta^{12} & \tilde\delta^{22} & \cdot \\
		\tilde\delta^{13} & \tilde\delta^{23} & \tilde\delta^{33}  
\ea\right)
\eea
with
\bea
\tilde\delta^{11}&=&1+\frac{3}{4}(\delta^{12}+\delta^{13})
-\frac{\sqrt{2}}{4}(\delta^{12}-\delta^{13})\\
\tilde\delta^{22}&=&1+\frac{3}{4}(\delta^{12}+\delta^{13})
+\frac{\sqrt{2}}{4}(\delta^{12}-\delta^{13}) \\
\tilde\delta^{33}&=&1+\frac{1}{2}(\delta^{12}+\delta^{13}) \\
\tilde\delta^{12}&=&\frac{1}{4}(\delta^{12}+\delta^{13}) \\
\tilde\delta^{13}&=&-\frac{1}{4}[(2-\sqrt{2})\delta^{12}-
(2+\sqrt{2})\delta^{13}] \\
\tilde\delta^{23}&=&-\frac{1}{4}[(2+\sqrt{2})\delta^{12}-
(2-\sqrt{2})\delta^{13}]
\eea
yielding the eigenvalues:
\bea
\tilde m^2_1,\quad 
\tilde m^2_2=\tilde m^2_1(1+2\delta^{12}),\quad
\tilde m^2_3=\tilde m^2_1(1+2\delta^{13}).
\eea
Now the mass splittings 
$\delta^{IJ}=(\tilde m^2_J-\tilde m^2_I)/(2\tilde m_1^2)$ are {\em not the
same as the off-diagonal mass insertions}.

\subsection{Charginos}

The chargino mass matrix, in the (charged wino, charged Higgsino) basis, is
\bea
{\bf X}=\left(\ba{cc} M_2 & \sqrt{2}M_W\sin\beta \\
	   	      \sqrt{2}M_W\cos\beta & \mu \ea\right)\; .
\eea
It can be diagonalized by two unitary matrices ${\bf U}$ and ${\bf V}$,
\bea
{\bf U}^* {\bf X} {\bf V}^{-1} = 
{\bf diag}(m_{\tilde\chi^\pm_1},m_{\tilde\chi^\pm_2})\; ,
\eea
where
\bea
m^2_{\tilde\chi^\pm_{1,2}}&=&\frac{1}{2}\Big[M^2_2+\mu^2+2M^2_W \nn\\
&\mp& \sqrt{(M^2_2-\mu^2)^2+4M^4_W\cos^2 2\beta+
4M^2_W(M^2_2+\mu^2+2M_2\mu\sin2\beta)} \Big]\; .
\eea
In order to get positive--mass eigenstates, one introduces two orthogonal
matrices ${\bf O_\pm}$,
\bea
{\bf U}&=&{\bf O_-}\; ,\\
{\bf V}&=&\left\{\ba{lcl} {\bf O_+}       &,& {\rm det}{\bf X}>0\; ,\\
		         \sigma_3{\bf O_+}&,& {\rm det}{\bf X}<0\; ,\ea\right.
\eea
where $\sigma_3$ is the usual Pauli matrix.

\subsection{Neutralinos}

The neutralino mass matrix, in the basis of the U(1) and SU(2) 
neutral gauginos and the two neutral Higgsinos $(\tilde B,\tilde W_3, \tilde
H^0_1, \tilde H^0_2)$, is the symmetric matrix:
\bea
{\bf Y}=\left(\ba{cccc} M_1 & \cdot & \cdot & \cdot \\
			0 & M_2 & \cdot & \cdot \\
			-M_Zs_W\cos\beta & M_Zc_W\cos\beta & 0 & \cdot \\
			M_Zs_W\sin\beta & -M_Zc_W\sin\beta & -\mu & 0
\ea\right)\; .
\eea
To simplify, we employ the unification constraint 
$M_1=\frac{5}{3}\tan^2\theta_W M_2$.

The matrix above can be numerically diagonalized by the unitary matrix $\N$,
\bea
\N^*{\bf Y}\N^{-1}=
{\bf diag}(m_{\tilde\chi^0_1},m_{\tilde\chi^0_2},
m_{\tilde\chi^0_3},m_{\tilde\chi^0_4})\; .
\eea

\subsection{Vertex couplings}

$\bullet$ Vertex $ig\gamma^\mu(g^L_{AB}P_L + g^R_{AB}P_R)$
[note that $g^L_{BA}=g^{L^*}_{AB}$, $g^R_{BA}=g^{R^*}_{AB}$]:
\bea
Z\bar{\tilde\chi}^-_A\tilde\chi^-_B:
&& g^L_{AB}= \frac{1}{c_W} O^{\prime L}_{AB};\quad 
O^{\prime L}_{AB}=\left(\frac{1}{2}-s^2_W\right) \U_{A2} \U_{B2}^* + 
                                   c^2_W \U_{A1} \U_{B1}^*\; ,
\\
&& g^R_{AB}= \frac{1}{c_W} O^{\prime R}_{AB};\quad 
O^{\prime R}_{AB}=\left(\frac{1}{2}-s^2_W\right) \V_{A2}^* \V_{B2} + 
                                   c^2_W \V_{A1}^* \V_{B1}\; .
\\
Z\bar{\tilde\chi}^0_A\tilde\chi^0_B:
&& g^L_{AB}= \frac{1}{c_W} O^{\prime\prime L}_{AB};\quad 
O^{\prime\prime L}_{AB}=\frac{1}{2}\left(\N_{A4} \N_{B4}^*-\N_{A3} \N_{B3}^*
\right)\; ,
\\
&& g^R_{AB}= \frac{1}{c_W} O^{\prime\prime R}_{AB};\quad
O^{\prime\prime R}_{AB}=-O^{\prime\prime L^*}_{AB}\; .
\eea
$\bullet$ Vertex $igG_{XY}(p_1+p_2)^\mu$
[note that $G_{YX}=G^*_{XY}$]:
\bea
Z\tilde\nu^\dagger_X\tilde\nu_Y:
&& G_{XY}= -\frac{1}{2c_W}\delta_{XY}\; ,
\\
Z\tilde\ell^\dagger_X\tilde\ell_Y:
&& G_{XY}=\frac{1}{c_W}\sum^3_{K=1} \left[
           \left(\frac{1}{2}-s^2_W\right)
	   \S^{\tilde\ell}_{XK} \S^{\tilde\ell^*}_{YK}
   - s^2_W \S^{\tilde\ell}_{X,K+3} \S^{\tilde\ell^*}_{Y,K+3}\right]\; .
\eea
$\bullet$ Vertex $ig(c^L_{IAX}P_L + c^R_{IAX}P_R)$:
\bea
\tilde\nu^\dagger_X\bar{\tilde\chi}^-_A\ell_I:
&& c^{L[C]}_{IAX}= -\V^*_{A1}\S^{\tilde\nu}_{XI}\; ,
\\
&& c^{R[C]}_{IAX}= \frac{m_{\ell_I}}{\sqrt{2}M_W\cos\beta}\U_{A2}
   \S^{\tilde\nu}_{XI}\; .
\\
\tilde\ell^\dagger_X\bar{\tilde\chi}^0_A\ell_I:
&& c^{L[N]}_{IAX}= \frac{1}{\sqrt{2}}(\tan\theta_W\N^*_{A1} + \N^*_{A2})\ 
              {\bf S}^{\tilde\ell}_{XI}	   
-\frac{m_{\ell_I}}{\sqrt{2}M_W\cos\beta}\N^*_{A3}\S^{\tilde\ell}_{X,I+3}\; ,
\hspace{1cm}
\\
&& c^{R[N]}_{IAX}=-\sqrt{2}\tan\theta_W\N_{A1} \S^{\tilde\ell}_{X,I+3} 
-\frac{m_{\ell_I}}{\sqrt{2}M_W\cos\beta}\N_{A3}\S^{\tilde\ell}_{XI}\; .
\eea

\section{The LFV decay $\ell_J\to\ell_I\gamma$ and $g-2$}

The general amplitude $\ell_J\to\ell_I\gamma$ at one loop reads
\bea
{\cal M}=-ie\frac{\alpha_W}{4\pi}\varepsilon^\mu_\gamma \bar u_{\ell_I}(p_2)
\frac{1}{m_{\ell_J}}\left[ (if^\gamma_M+f^\gamma_E\gamma_5)
\sigma_{\mu\nu}q^\nu\right]
u_{\ell_J}(p_1)\; .
\eea
In the literature one finds often the notation:
\bea
 \frac{\alpha_W}{4\pi}f^\gamma_M=\frac{m^2_{\ell_J}}{2}(A_2^L+A_2^R)\; ,\quad
i\frac{\alpha_W}{4\pi}f^\gamma_E=\frac{m^2_{\ell_J}}{2}(A_2^L-A_2^R)\; .
\eea
For equal leptons, the anomalous magnetic dipole moment of $\ell$ is
\bea
\delta a_\ell=\frac{g_\ell-2}{2}=\frac{\alpha_W}{4\pi} f^\gamma_M
=\frac{m^2_{\ell}}{2}(A_2^L+A_2^R)\; .
\eea

\noindent
The width of $\ell_J\to\ell_I\gamma$ is
\bea
\Gamma(\ell_J\to\ell_I\gamma)
=\frac{\alpha\alpha_W^2}{32\pi^2}m_{\ell_J}
 (|f^\gamma_M|^2+|f^\gamma_E|^2)
=\frac{\alpha}{4}m_{\ell_J}^5(|A_2^L|^2+|A_2^R|^2)\; .
\eea
Since the width 
$\Gamma(\ell_J\to \ell_I\nu_J\bar\nu_I)=
\displaystyle\frac{G^2_Fm^5_{\ell_J}}{192\pi^3}$ and
$G_F=\displaystyle\frac{\pi\alpha_W}{\sqrt{2}M^2_W}$,
one has
\bea
\frac{{\rm BR}(\ell_J\to \ell_I\gamma)}
{{\rm BR}(\ell_J\to \ell_I\nu_J\bar\nu_I)}=
\frac{12\alpha}{\pi}\frac{M^4_W}{m^4_{\ell_J}}
(|f^\gamma_M|^2+|f^\gamma_E|^2)=
\frac{48\pi^3\alpha}{G^2_F}(|A^L_2|^2+|A^R_2|^2)\; ,
\eea
where ${\rm BR}(\ell_J\to \ell_I\nu_J\bar\nu_I)=1/0.17/0.17$ for 
$\ell_J\ell_I=\mu e/\tau e/\tau\mu$, respectively.

The SUSY contributions to the form factors are the following.

\noindent
$\bullet$ Diagram of type A: Chargino--Chargino--Sneutrino 
[$x_{AX}=m^2_{\tilde\chi^\pm_A}/m^2_{\tilde\nu_X}$]:
\bea
\left.\frac{f^\gamma_M}{m_{\ell_J}}\right|_{\tilde\chi^\pm}=
\sum_{A=1}^{2}\sum_{X=1}^{3}\left[
 \frac{m_{\ell_J}}{m^2_{\tilde\nu_X}}c^{L[C]*}_{IAX}c^{L[C]}_{JAX} 
F_1(x_{AX})+\frac{m_{\tilde\chi^\pm_A}}{m^2_{\tilde\nu_X}}
                                     c^{L[C]*}_{IAX}c^{R[C]}_{JAX} 
F_2(x_{AX})+L\leftrightarrow R\right]\; ,
\label{fmc}\\
\left.i\frac{f^\gamma_E}{m_{\ell_J}}\right|_{\tilde\chi^\pm}=
\sum_{A=1}^{2}\sum_{X=1}^{3}\left[
 \frac{m_{\ell_J}}{m^2_{\tilde\nu_X}}c^{L[C]*}_{IAX}c^{L[C]}_{JAX} 
F_1(x_{AX})+\frac{m_{\tilde\chi^\pm_A}}{m^2_{\tilde\nu_X}}
                                     c^{L[C]*}_{IAX}c^{R[C]}_{JAX} 
F_2(x_{AX})-L\leftrightarrow R\right]\; .
\label{fec}
\eea
$\bullet$ Diagram of type B: Slepton--Slepton--Neutralino 
[$x_{AX}^0=m^2_{\tilde\chi^0_A}/m^2_{\tilde\ell_X}$]:
\bea
\left.\frac{f^\gamma_M}{m_{\ell_J}}\right|_{\tilde\chi^0}=
\sum_{A=1}^{4}\sum_{X=1}^{6}\left[
 \frac{m_{\ell_J}}{m^2_{\tilde\ell_X}}c^{L[N]*}_{IAX}c^{L[N]}_{JAX} 
F_3(x_{AX}^0)+\frac{m_{\tilde\chi^0_A}}{m^2_{\tilde\ell_X}}
                                      c^{L[N]*}_{IAX}c^{R[N]}_{JAX}
F_4(x_{AX}^0)+L\leftrightarrow R\right]\; ,
\label{fmn}\\
\left.i\frac{f^\gamma_E}{m_{\ell_J}}\right|_{\tilde\chi^0}=
\sum_{A=1}^{4}\sum_{X=1}^{6}\left[
 \frac{m_{\ell_J}}{m^2_{\tilde\ell_X}}c^{L[N]*}_{IAX}c^{L[N]}_{JAX} 
F_3(x_{AX}^0)+\frac{m_{\tilde\chi^0_A}}{m^2_{\tilde\ell_X}}
                                      c^{L[N]*}_{IAX}c^{R[N]}_{JAX}
F_4(x_{AX}^0)-L\leftrightarrow R\right]\; ,
\label{fen}
\eea
where
\bea
F_1(x)&=&\frac{2+3x-6x^2+x^3+6x\ln x}{6(1-x)^4}\; ,
\\
F_2(x)&=&\frac{-3+4x-x^2-2\ln x}{2(1-x)^3}\; ,
\\
F_3(x)&=&-\frac{1-6x+3x^2+2x^3-6x^2\ln x}{6(1-x)^4}=-xF_1(1/x)\; ,
\\
F_4(x)&=&-\frac{1-x^2+2x\ln x}{2(1-x)^3}\; .
\eea
These functions are combinations of 3--point tensor integrals, in
agreement with \cite{Hollik:1999vz}:
\bea
F_1(x_{AX})/m^2_{\tilde\nu_X}&=&2\ [C_{11}+C_{21}+C_{23}]
(0,0,0;m_{\tilde\nu_X},m_{\tilde\chi^\pm_A},m_{\tilde\chi^\pm_A})\; , \\
F_2(x_{AX})/m^2_{\tilde\nu_X} &=&2\ C_{11}(0,0,0;m_{\tilde\nu_X},
m_{\tilde\chi^\pm_A},m_{\tilde\chi^\pm_A})\; , \\
F_3(x_{AX}^0)/m^2_{\tilde\ell_X}&=&-2\ [C_{11}+C_{21}+C_{23}](0,0,0;
m_{\tilde\chi^0_A},m_{\tilde\ell_X},m_{\tilde\ell_X})\; , \\
F_4(x_{AX}^0)/m^2_{\tilde\ell_X}&=&[C_{0}+C_{11}+C_{12}](0,0,0;
m_{\tilde\chi^0_A},m_{\tilde\ell_X},m_{\tilde\ell_X})\; .
\eea
Note that the dipole form factors (\ref{fmc}--\ref{fen}) 
are proportional to 
a fermion mass. The chirality flip takes place in the external
fermion lines, for the terms proportional to $LL$ and $RR$ mixings
 and in the internal fermion lines (charginos or neutralinos), 
for the terms proportional to the $LR$ mixing.

The branching ratio $\ell_J\to \ell_I\gamma$ reads
\bea
{\rm BR}(\ell_J\to \ell_I\gamma)&=&
{\rm BR}(\ell_J\to \ell_I\nu_J\bar\nu_I)\times
\frac{12\pi\alpha\alpha_W^2}{G^2_F}\nonumber\\
&&\hspace*{-3cm}\times
\Bigg\{\left| \sum_{AX}\frac{1}{m_{\tilde \nu_X}^2}
\left( c^{L[C]^*}_{IAX}c^{L[C]}_{JAX} F_1(x_{AX}) +
\frac{m_{\tilde\chi^\pm_A}}{m_{\ell_J}} c^{L[C]^*}_{IAX}c^{R[C]}_{JAX} 
F_2(x_{AX})
\right)\right.  \nonumber\\&&\hspace*{-2.1cm}
\left. \sum_{AX}\frac{1}{m_{\tilde\ell_X}^2}
\left( c^{L[N]^*}_{IAX}c^{L[N]}_{JAX} F_3(x_{AX}^0) +
\frac{m_{\tilde\chi^0_A}}{m_{\ell_J}} c^{L[N]^*}_{IAX}c^{R[N]}_{JAX}
 F_4(x_{AX}^0)\right)\right|^2+ L\leftrightarrow R\Bigg\}\; .
\hspace{8mm}
\eea


\end{document}